\documentclass[letterpaper]{article} 
\usepackage{aaai25}  
\usepackage{times}  
\usepackage{helvet}  
\usepackage{courier}  
\usepackage[hyphens]{url}  
\usepackage{graphicx} 
\usepackage{natbib}  
\usepackage{caption} 
\usepackage{algorithm}
\usepackage{algorithmic}

\usepackage{newfloat}
\usepackage{listings}
\DeclareCaptionStyle{ruled}{labelfont=normalfont,labelsep=colon,strut=off} 
\floatstyle{ruled}
\newfloat{listing}{tb}{lst}{}
\floatname{listing}{Listing}
\usepackage[utf8]{inputenc} 
\usepackage{url}            
\usepackage{booktabs}       
\usepackage{amsfonts}       
\usepackage{nicefrac}       
\usepackage{microtype}      

\usepackage{epsfig}
\usepackage{amsmath}
\usepackage{amssymb}
\usepackage{makecell}
\usepackage{multirow}

\usepackage{tikz}
\usepackage{subcaption}
\usepackage{dsfont}
\usepackage{filecontents}
\usepackage{textcomp}
\usepackage{mathrsfs}

\usepackage{lipsum}

\usepackage{endnotes}

\usepackage{amsthm}

\theoremstyle{plain}
\newtheorem{theorem}{Theorem}
\newtheorem{proposition}[theorem]{Proposition}
\usepackage[font=bf]{caption}
\usepackage{url}
\usepackage{xspace} 
\usepackage{epstopdf}
\usepackage{bm}
\usepackage{lineno}
\usepackage{rotating}

\newcommand{\name}{\text{TrojanDec}}
\newcommand{\myparatight}[1]{\smallskip\noindent{\bf {#1}:}~}
\newcommand{\myproof}[1]{\smallskip\noindent{\textit {#1}.}~}

\newcommand{\argmin}{\operatornamewithlimits{argmin}}

\newcommand\CR[1]{\textcolor{black}{#1}}

\title{{\name}: Data-free Detection of Trojan Inputs in Self-supervised Learning}
\author{
    Yupei Liu, Yanting Wang, Jinyuan Jia
}
\affiliations{
    The Pennsylvania State University\\

    \{yzl6415, ykw5450, jinyuan\}@psu.edu
}

\usepackage{bibentry}

\begin{document}

\maketitle

\begin{abstract}
An image encoder pre-trained by self-supervised learning can be used as a general-purpose feature extractor to build downstream classifiers for various downstream tasks. However, many studies showed that an attacker can embed a trojan into an encoder such that multiple downstream classifiers built based on the trojaned encoder simultaneously inherit the trojan behavior. In this work, we propose {\name}, the first data-free method to identify and recover a test input embedded with a trigger. Given a (trojaned or clean) encoder and a test input, {\name} first predicts whether the test input is trojaned. If not, the test input is processed in a normal way to maintain the utility. Otherwise, the test input will be further restored to remove the trigger. Our extensive evaluation shows that {\name} can effectively identify the trojan (if any) from a given test input and recover it under state-of-the-art trojan attacks. We further demonstrate by experiments that our {\name} outperforms the state-of-the-art defenses. 

\end{abstract}

\section{Introduction}
\label{sec:introduction}
Traditional transfer/supervised learning trains a feature extractor using a labeled training dataset, which incurs large costs and human effort to annotate the training dataset. Moreover, it cannot leverage a large amount of unlabeled data that can be collected from the Internet.
Self-supervised learning~\cite{devlin2018bert,hadsell2006dimensionality,he2020momentum,chen2020simple,hjelm2018learning,grill2020bootstrap} has been designed to address those challenges. In particular, a model provider can use self-supervised learning to pre-train an encoder using a large set of unlabeled data (called \emph{pre-training dataset}). This paradigm can significantly save the labeling effort and thus enables the model provider to significantly increase the size of its pre-training dataset. For example, CLIP~\cite{radford2021learning} was pre-trained by OpenAI on 400 million (image, text) pairs collected from the Internet. The model provider can monetize its encoder by deploying it as a cloud service via providing an API to users~\cite{liu2022stolenencoder,liu2024formal}. Suppose a user has some training/test inputs for a downstream task. The user can query the API to obtain their feature vectors and then use those  feature vectors to train/test a downstream classifier for its downstream task. 

Despite the success of self-supervised learning, many existing studies show that it is vulnerable to trojan attacks~\cite{jia2021badencoder,saha2022backdoor,carlini2021poisoning}. Specifically, an adversary can inject a trojan to self-supervised learning encoders by either poisoning the pre-training dataset~\cite{saha2022backdoor,liu2022poisonencoder,carlini2021poisoning} or directly manipulating pre-trained encoders' parameters~\cite{jia2021badencoder}. The trojaned encoder will produce normal feature vectors for clean inputs, but output feature vectors similar to the feature vector of an attacker-chosen reference object for any test inputs with the attacker-chosen trigger. Furthermore, the trojan behavior will be conveyed to downstream classifiers built upon the trojaned encoder, leading those classifiers to predict the test inputs with the trigger to the ground-truth label (called \emph{target label}) of the attacker-chosen reference object. 

Many defenses~\cite{tran2018spectral,chen2018detecting,wang2019neural,xu2019detecting,chen2019deepinspect,liu2019abs,gao2019strip,doan2020februus,li2021neural} were proposed to mitigate trojan attacks to machine learning models. Depending on which stage those defenses are used, we categorize them into \emph{training-phase defenses}~\cite{tran2018spectral,chen2018detecting,wang2019neural,xu2019detecting,chen2019deepinspect,liu2019abs,li2021neural} and \emph{testing-phase defenses}~\cite{doan2020februus,li2021neural}. A training-phase defense either trains a robust model on a poisoned training dataset or detects/removes the trojan in a trained model. As a result, they require access to the training dataset or the model parameters of the model. In other words, they are not applicable when a defender only has black-box access to a pre-trained model. In our work, we consider a defender (i.e., user) has black-box access to an encoder (we discuss more details in our threat model) and thus those defenses are not applicable in general. Moreover, when extending to self-supervised learning, some of these defenses are ineffective even if we assume the defender has white-box access to the encoder as shown in previous studies~\cite{jia2021badencoder} (we also confirm this in our experiments). Testing-phase defenses~\cite{doan2020februus,li2021neural} aims to detect whether a test input is trojaned.  When extended to self-supervised learning, those defenses are either ineffective or require a defender to have a clean validation dataset, which is less practical in the real world. In our work, we relax such an assumption by proposing a data-free defense. Moreover, our comparison results show that our defense can achieve comparable performance by giving those defenses an advantage, i.e., assuming the defender has a clean validation dataset.

\myparatight{Our work} In this work, we propose {\name}, the first framework to identify and restore a trojaned test image in the self-supervised learning context. Our framework consists of three main components: 1) metadata extraction, 2) trojan detection, and 3) image restoration. In the first component, we extract the key metadata from a test image, which is critical in detecting if the image consists of a trigger. In the second component, we perform a data-free statistical analysis on the metadata to identify if the corresponding image is trojaned or not. In the third component, if the image is detected as a trojaned example in the previous part, we further restore it using a diffusion model. 

To show the effectiveness of our {\name}, we conduct extensive experiments on multiple pre-training datasets and downstream tasks under state-of-the-art trojan attacks~\cite{jia2021badencoder,saha2022backdoor,liu2022poisonencoder,zhang2022corruptencoder} to self-supervised learning. Our results some that our defense is consistently effective under those attacks. We further generalize several representative backdoor attacks~\cite{chen2017targeted,salem2020dynamic} designed for supervised learning to the self-supervised learning context as the adaptive attacks and show the effectiveness of our method in defending against them. To study the impact of hyperameters, we perform a comprehensive ablation study in our evaluation.  Finally, we compare our defense with existing defenses~\cite{gao2019strip,doan2020februus,ma2022beatrix} against trojan attacks. Our results demonstrate that {\name} outperforms all of them. 

To summarize, we make the following contributions: 
\begin{itemize}
    \item We propose the first generic data-free trojaned test input detection and restoration framework for self-supervised learning.
    
    \item We demonstrate that our framework does not rely on any clean data or any prior knowledge to the pre-training or downstream dataset. 
    
    \item We conduct extensive experiments to evaluate our {\name} and show that it is effective in defending against various types of trojan attacks and outperforms existing defense methods. 
\end{itemize}
\section{Related Work}
\label{sec:relatedwork}

\subsection{Trojan Attacks}

\myparatight{Trojan attacks to self-supervised learning}
Trojan attacks~\cite{carlini2021poisoning,saha2022backdoor,jia2021badencoder,li2023embarrassingly,zhang2022corruptencoder,wu2023efficient,jha2023labelpoisoningneed,boberirizar2023architectural,li2024on} to self-supervised learning aim to produce a trojan encoder such that it outputs normal feature vectors for clean test inputs, while generating feature vectors that are similar to the feature vector of an attacker-chosen reference object (called target object) for test inputs embedded with a trigger. As a result, a downstream classifier built upon such a trojan encoder for a downstream task inherits such trojan behaviors, i.e., the downstream classifier predicts the same class as the reference object for any test input embedded with the trigger. 
Depending on how an attacker injects a trojan into an encoder, those attacks can be divided into two types. The first type of attack is based on poisoning the pre-training dataset used to pre-train the encoder~\cite{saha2022backdoor,liu2022poisonencoder,carlini2021poisoning,zhang2022corruptencoder}. For instance,~\citet{saha2022backdoor} proposed to create  poisoned examples by placing  a trigger near the target object. The other type of trojan attack is based on directly manipulating the parameters of a pre-trained encoder~\cite{jia2021badencoder}. For example, BadEncoder~\cite{jia2021badencoder} leverages a surrogate dataset to fine-tune a pre-trained encoder to maximize the similarity between the features vectors of a trojan image and the attacker-chosen reference object. This kind of attack is usually more powerful than the previous attack as an attacker can arbitrarily manipulate the parameters. In our work, we consider both types of attacks. 

\myparatight{Trojan attacks to supervised learning}
We note that backdoor attacks have also been studied for supervised learning~\cite{gu2017badnets,chen2017targeted,liutrojaning2018,liu2020reflection,li2021invisible,zeng2021rethinking}. In particular, those studies try to design different triggers to make the attak more effective and stealthy.
For instance, the dynamic trojan attack~\cite{salem2020dynamic} changes the pattern and the location of the trigger frequently to avoid being detected. The blending attack~\cite{chen2017targeted} mixes the trigger pattern into the original input by using the background noise as the trigger. 
Those trojan attacks are designed for classifiers instead of self-supervised learning encoders. In our experiments, we will generalize them to self-supervised learning and our results show that our defense is effective against those attacks.

\begin{figure}[!t]
	 \centering
{\includegraphics[width=0.47\textwidth]{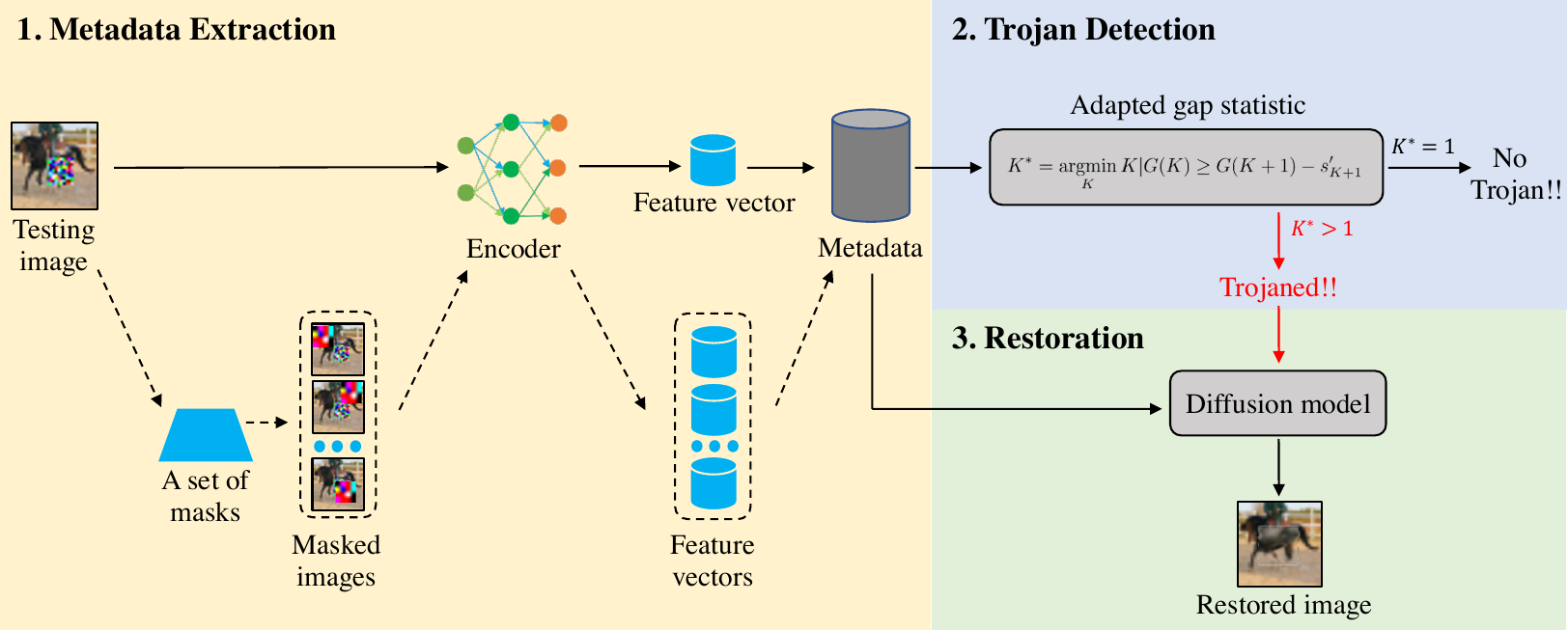}}
\caption{An overview of our {\name}. } 
\label{fig:overview}
\end{figure}

\subsection{Defending against Trojan Attacks} 
There are many existing works that focus on defending against trojan attacks~\cite{tran2018spectral,chen2018detecting,wang2019neural,xu2019detecting,chen2019deepinspect,liu2019abs,gao2019strip,doan2020februus,li2021neural,jia2020certified,cheng2023beagle,huang2022decoupling}. Most of them are designed for supervised learning. Depending on which phase those defenses are applied, they can be categorized into two genres: training-phase defense  and testing-phase defense. 

\myparatight{Training-phase defenses}
Training-phase defenses~\cite{chen2018detecting,wang2019neural,kolouri2020litmus,huang2022decoupling,zheng2022lipschitzness,tejankar2023defending,feng2023detect,zheng2024sslcleanse,wang2024mmdb} take two directions. In the first direction, they aim to train a robust model on a poisoned pre-training dataset. They are not applicable when an attacker implants a trojan into a model by directly manipulating its parameters~\cite{jia2021badencoder}. In the second direction, they aim to detect and remove the trojan for a given model. For instance, Neural Cleanse views each output class in a classifier as a potential target class and reverses engineer a trojan trigger for it~\cite{wang2019neural}. Then, it uses outlier detection statistics to determine if the classifier is trojan. The defenses designed in this direction are mostly tailored for supervised learning. Moreover, many of those defenses~\cite{wang2019neural,zheng2024sslcleanse} require accessing the parameters of the model, which make them not applicable when the defender does not have such access, e.g., the model is deployed as a cloud service and the defender can only query it. 
Additionally, existing studies~\cite{jia2021badencoder} showed that some defenses~\cite{wang2019neural,xu2019detecting} in this direction are ineffective when extended to self-supervised learning. In general, our testing-phase defense is complementary to training-phase defense. 

\myparatight{Testing-phase defenses}
Testing-phase defenses~\cite{gao2019strip,doan2020februus,ma2022beatrix,xi2024mdp} aim to detect or recover a test input that potentially contains a trigger. Most of these defenses require the defender to have some clean data. In our work, we consider a more realistic scenario where the defender does not have any clean data. Our experimental results show that our defense can achieve comparable performance with those defenses by giving them the advantage (i.e., we assume the defender has some clean data for those defenses). 

\CR{For space reasons, more discussion on related work can be found in our Appendix.}

\section{Threat Model} 
\label{problem}

\myparatight{Attacker's goal} To perform trojan attacks to
self-supervised learning, an attacker selects one (or more) target classes in one (or more) downstream tasks.
The attacker then chooses a trigger for each target class. In particular, the attacker aims at crafting a trojaned encoder to achieve two goals. Firstly, the encoder generates normal feature vectors for clean test inputs (i.e., the inputs without triggers). This means that the functionality of the trojaned encoder is maintained for clean inputs, which makes the attack more stealthy and harder to be noticed. Secondly, when the trigger presents in a test input, the trojaned encoder will produce a feature vector that lies in the target class associated with that trigger. As a result, a downstream classier built on the trojan encoder for a downstream task will predict the target class for any input with a backdoor trigger.

\myparatight{Attacker's background knowledge and capability} We consider both data-poisoning based attacks~\cite{saha2022backdoor,carlini2021poisoning} and model-poisoning based attacks~\cite{jia2021badencoder}. For data-poisoning based attacks, we consider an attacker can inject poisoned training inputs to the pre-training dataset used to pre-train an encoder. For model-poisoning based attacks, we consider an attacker can arbitrarily manipulate the parameters of an encoder. In general, model-poisoning based attacks are more effective than data-poisoning based attacks as they make a stronger assumption on the attacker.

\myparatight{Defender's goal} We consider the defender as the downstream user of the deployed self-supervised learning encoder. Given a test input, the defender's goal is to detect whether it contains a trigger. If this input is identified as trojaned, the defender wants to further remove the trigger from the input.

\myparatight{Defender's background knowledge and capability} We consider a very weak defender. As mentioned, the defender is a downstream user of the pre-trained encoder. In this scenario, the defender only has the "black-box" access to the encoder, which means the following: 1) the defender can only query the encoder by images and receive the produced feature vectors and 2) the defender does not have any knowledge to the encoder, including the pre-training dataset information, the encoder architecture, the encoder parameters, etc.

\section{Our {\name}} 
\label{method}

\begin{algorithm}[tb]
   \caption{Metadata Extraction}
   \label{alg:detection}
\begin{algorithmic}[1]
   \STATE {\bfseries Input:} $\mathbf{x}$ (test input), $f$ (encoder), $k$ (mask size), $s$ (step size), and $t$ (image size).
   \STATE {\bfseries Output:}  $\mathcal{S}$ (the metadata of $\mathbf{x}$) \\
    
   \STATE $\mathcal{M} \gets \textit{MaskSetGeneration}(k, s, t)$ \\
   \STATE $\mathcal{S} \gets \emptyset$ \\

   \FOR{$(\mathbf{m}, \mathbf{p}) \textrm{  in  } \mathcal{M}$}
  \STATE $\mathbf{x}_{\mathbf{m}, \mathbf{p}} \gets \mathbf{m} \cdot \mathbf{x} + (1 - \mathbf{m}) \cdot \mathbf{p}$ \\
   \STATE $\textit{d} \gets sim(f(\mathbf{x}_{\mathbf{m}, \mathbf{p}}), f(\mathbf{x}))$ \\
   \STATE $\mathcal{S}$.$\textrm{append}(\textit{d})$ \\
   \ENDFOR
   
   \STATE \textbf{return} $S$
\end{algorithmic}
\end{algorithm}

We present {\name} as an end-to-end framework for identifying and eliminating backdoors from test images. As depicted in Figure~\ref{fig:overview}, our framework consists of 3 key components: the extraction of metadata from a test input, backdoor detection based on metadata, and test input restoration. \CR{We also assume the attacker uses a patch-based trojan trigger, since patch-based triggers are prevalent in real-world scenarios as they are practical and easy to implement physically.
} Next, we delve into details of these components.

\subsection{Metadata Extraction}
\label{method:sub1}

There are 3 steps to extract metadata from a test image. First, we generate a set of masks and use them to mask the test image to create a set of masked images. Then, we query the encoder $f$ (trojaned or not) to obtain feature vectors of masked images and the original test image. Finally, we compute the similarity between the feature vectors of each masked image and the original test image to derive the metadata. 

For simplicity, we use $(\mathbf{m}, \mathbf{p})$ to denote a patch-based mask, where $\mathbf{m}$ is a binary offset and $\mathbf{p}$ is the mask pattern. The binary offset $\mathbf{m}$ controls two characteristics of the mask: location and size. Specifically, we denote the upper-left coordinate of a mask as $(a, b)$ and the mask size as $k$. Given the location $(a, b)$ and mask size $k$, we define the binary offset $\mathbf{m}$ at the coordinate $(i, j)$ as follows: 

\begin{align}
\label{offset_definition}
\mathbf{m}_{i, j} = 
  \left\{ 
    \begin{array}{ c l }
        0 & \quad \textrm{if } i \in [a,a+k), j \in [b,b+k) \\
        1 & \quad \textrm{otherwise}
    \end{array}
  \right.
\end{align}

The mask pattern $\mathbf{p}$ controls the pixel values of the mask. In particular, when the mask is applied to a test image, for a specific coordinate $(i, j)$, $\mathbf{m}_{i, j} = 1$ means it keeps the original pixel values of the image at $(i, j)$ and $\mathbf{m}_{i, j} = 0$ means it masks this coordinate out (i.e., reset the pixel value of the image at $(i, j)$ to $\mathbf{p}_{i, j}$).  Given a test image $\mathbf{x}$, we obtain the masked image $\mathbf{x}_{\mathbf{m}, \mathbf{p}}$ by applying $(\mathbf{m}, \mathbf{p})$ to $\mathbf{x}$ as follows: 

\begin{align}
\label{apply_mask}
  \mathbf{x}_{\mathbf{m}, \mathbf{p}} = \mathbf{m} \cdot \mathbf{x} + (1 - \mathbf{m}) \cdot \mathbf{p}
\end{align}

We notice that pre-defining $\mathbf{p}$ to a fixed pattern can be dangerous, as the attacker can exploit this fact and set the trojan trigger to have the same pattern as $\mathbf{p}$. Thus, for each mask, we randomly generate a unique pattern such that for each coordinate $(i, j)$, $\mathbf{p}_{i, j}$ is randomly sampled from 0 to 255. Formally, we have the following proposition. 

\begin{proposition}
\label{prop_1}
If the adversary sets the trojan trigger to be $\mathbf{e}$ whose height and width are $e_h$ and $e_w$, while defender has a mask $(\mathbf{m}, \mathbf{p})$ such that the pattern $\mathbf{p}$ is randomly generated, the probability of the existence of a part of the mask pattern such that $\ell_1$ distance of this part of the mask pattern and the trojan trigger greater than $\beta$ is no smaller than $1-\frac{(2\beta)^{T}}{T!}$, where $T=e_h \cdot e_w$.
\end{proposition}

\myproof{Proof} See Appendix. 

Following this rule, we can generate a set of masks, namely $\mathcal{M}$. Next, for each mask $(\mathbf{m}, \mathbf{p})$ in $\mathcal{M}$, we apply it to the given input to obtain a masked image $\mathbf{x}_{\mathbf{m}, \mathbf{p}}$. We then use each of them to query $f$. For a masked image $\mathbf{x}_{\mathbf{m}, \mathbf{p}}$, we will receive its corresponding feature vector $f(\mathbf{x}_{\mathbf{m}, \mathbf{p}})$. We also query $f$ using the original test image to receive $f(\mathbf{x})$. Finally, we calculate the cosine similarity between the features of $\mathbf{x}_{\mathbf{m}, \mathbf{p}}$ and $\mathbf{x}$, i.e., $sim(f(\mathbf{x}_{\mathbf{m}, \mathbf{p}}), f(\mathbf{x}))$. After this process, we will obtain the metadata $\mathcal{S}$ of $\mathbf{x}$, where $\mathcal{S}$ consists of a set of similarities between masked images and $\mathbf{x}$. 

Algorithm 2 summarizes how {\name} extracts the metadata from a test image $\mathbf{x}$. Algorithm 2 in the Appendix presents details of how {\name} generates the mask set $\mathcal{M}$. The helper function $\textit{CreateMask}(a, b, k)$  creates a mask $(\mathbf{m}, \mathbf{p})$ with the upper-left coordinate being $(a, b)$, mask size being $k$, offset values being defined in Equation~\ref{offset_definition}, and mask pattern $\mathbf{p}$ being randomly generated. An example of such mask with randomly generated pattern is in Figure~\ref{fig:example_masked}.

\subsection{Trojan Detection}
\label{method:sub2}
\begin{figure}[!t]
	 \centering
{\includegraphics[width=0.33\textwidth]{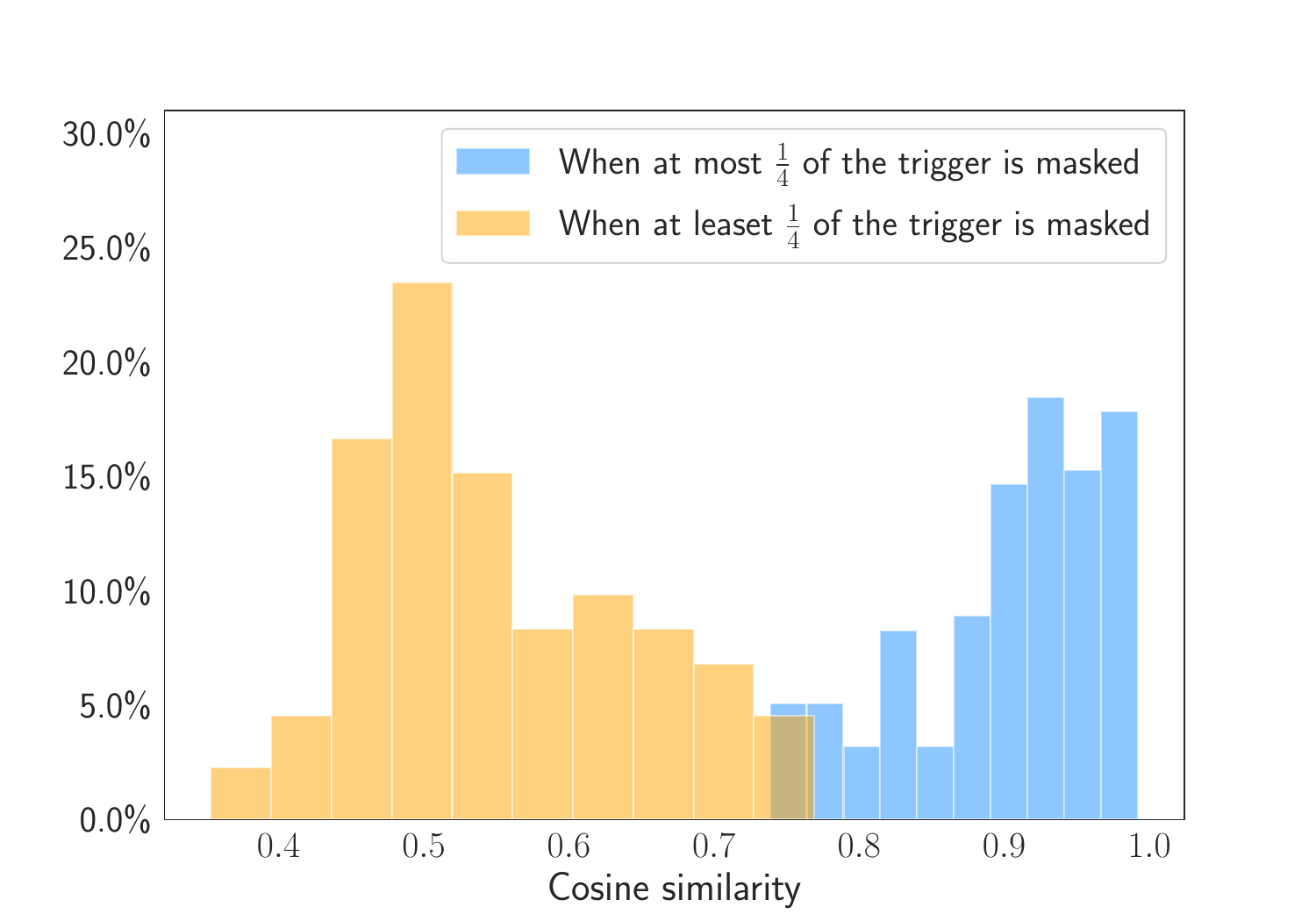}}
\caption{Cosine similarities of the feature vectors of masked images and the feature vector of the original trojaned test image. The trojaned encoder is pre-trained on CIFAR10 and the downstream dataset is STL10. } 
\label{dist}
\end{figure}

In this section, we present our technique for detecting whether a test image is trojaned or not, based on its metadata. The key intuition of our {\name} is that the set of similarities in $\mathcal{S}$ can be divided into two clusters for a trojaned test image, whereas there is only a single cluster for a clean test image. Figure~\ref{dist} shows the distribution of $\mathcal{S}$ of a test image sampled from STL10 dataset. When the trigger is partially covered by the mask, the trojan information of this test image is broken, leading the feature vectors of these masked images being dissimilar to the original trojaned image. In Figure~\ref{dist}, when at least one fourth of the trojan trigger is occluded by the mask, the similarities between the feature vectors generated by $f$ for masked images and the original image are much lower (i.e., most of them are less than 0.7), whereas the majority of the similarities in the other cluster are greater than 0.9. Therefore, by the number of clusters $\mathcal{S}$ can be classified to, we can tell if the test image is trojaned. 

To determine the number of clusters the similarity scores in $\mathcal{S}$ belongs to, we leverage gap-statistic~\cite{tibshirani2001gap} to reach the goal. Suppose $\mathcal{S}$ can be divided into $K$ clusters, where $K=1$ or $2$. We first use \textit{K-means clustering} method~\cite{hartigan1979kmeans} to divide $\mathcal{S}$ into $K$ clusters. Then, we derive the within-cluster dispersion $W_K$, which measures the variability of the data points within each cluster. We can use $W_K$ to  calculate the gap statistics. To reach the goal, we first generate $B$ synthetic datasets based on a uniform distribution, then use K-means clustering to divide each of them into $K$ groups, and calculate the within-cluster dispersion $W_{Kb}^*$ for each of them, where $b=1, 2, \cdots, B$. The gap statistic for $K$ is defined as follows: 
\begin{align}
\label{eq:gap_statistic}
  G(K) = \frac{1}{B}\sum_{b=1}^{B}[\log{(W_{Kb}^*)} - \log{(W_K)}]
\end{align}
Then, we calculate the standard deviation of $\log{W_{Kb}^*}$, which is denoted as $s_K$.
In particular, based on the definition of the standard deviation, we have  $s_K = [\frac{1}{B}\sum_{b=1}^{B}(\log{(W_{Kb}^*)} -\bar{l})^2]^{\frac{1}{2}}$, where $\bar{l} = \frac{1}{B}\sum_{b=1}^{B}\log{(W_{Kb}^*)}$. The standard deviation is further normalized by the number of synthetic datasets $B$. Specifically, we have $s'_K = s_K\sqrt{1 + \frac{1}{B}}$. Finally, we derive the optimal number of clusters of $\mathcal{S}$ via solving the following optimization problem:
\begin{align}
\label{eq:s_k}
K^* = \argmin_K K 
  \text{ s.t. } G(K) \geq G(K+1) - s'_{K+1}
\end{align}

Given a test image $\mathbf{x}$ and its metadata $\mathcal{S}$, we derive the optimal number of clusters $K^*$. If $K^*=1$, it means $\mathbf{x}$ is clean. Otherwise, we predict $\mathbf{x}$ as a trojaned image.

\subsection{Image Restoration}
\label{method:sub3}

If a test input is identified as trojaned, we aim to remove the trojan trigger to restore the image. Given a test image $\mathbf{x}$ that is detected as a trojaned input, we use its masked version $\mathbf{x}_{\mathbf{m}^i, \mathbf{p}^i}$ as the prototype for restoration, where $i = \argmin_i sim(\mathbf{x}_{\mathbf{m}^i, \mathbf{p}^i}, \mathbf{x})$ and $sim$ is cosine similarity. Our intuition is that since $\mathbf{x}$ is detected as a trojaned input, $\mathbf{x}_{\mathbf{m}^i, \mathbf{p}^i}$ is the masked image that most likely to have the entire trojan trigger covered by the mask. A naive way is to directly use the prototype $\mathbf{x}_{\mathbf{m}^i, \mathbf{p}^i}$ as the ``restored'' image. However, as existing works reveal, this will substantially demolish the image quality and may change the semantic meaning of the original image~\cite{doan2020februus}. Thus, we use DDNM (denoising diffusion null-space model)~\cite{wang2022ddnm} to restore the masked image. Given an image $\mathbf{x}_0$, DDNM adds Gaussian noise in a step-by-step manner. In particular, the noisy image in the $t^\textit{th}$ step is $\mathbf{x}_t=\sqrt{\bar{\alpha}_t}\mathbf{x}_0 + \epsilon\sqrt{1-\bar{\alpha}_t}$, where $\bar{\alpha}_t$ is a pre-defined scalar constant and $\epsilon \sim \mathcal{N}(0, \mathbf{I})$ represents the zero-mean Gaussian noise. The goal of DDNM is to train a denoise model to estimate the noise added to $\mathbf{x}_0$ based on $\mathbf{x}_t$. As a result, the denoise model can be used to recover noisy images. 

\begin{figure}[!t]
	 \centering
\subfloat[]{\includegraphics[width=0.05\textwidth]{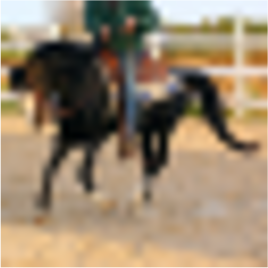}\label{fig:example_original}}
\qquad
\subfloat[]{\includegraphics[width=0.05\textwidth]{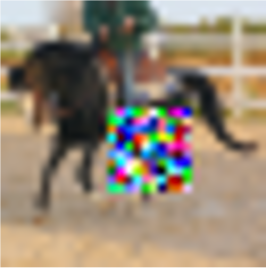}\label{fig:example_trojaned}}
\qquad
\subfloat[]{\includegraphics[width=0.05\textwidth]{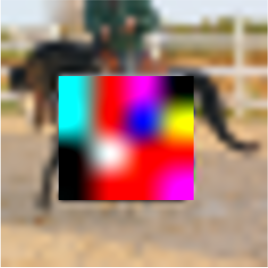}\label{fig:example_masked}}
\qquad
\subfloat[]{\includegraphics[width=0.05\textwidth]{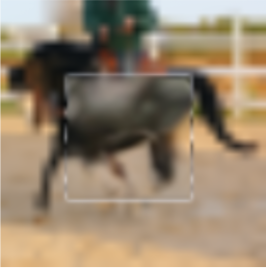}\label{fig:example_restored}}
\caption{Example of an image from the STL10 dataset recovered from its masked prototype: (a) original, (b) trojaned, (c) masked, and (d) restored.}
\label{fig:example}
\end{figure}

DDNM can be utilized to recover the masked image, since it can be used as a zero-shot solver for linear image restoration problems by refining only the masked area in the reverse diffusion process. We leverage the denoise model to recover the masked image. Based on the definition, we have $\mathbf{x}_{\mathbf{m}^i, \mathbf{p}^i} = \mathbf{m}^i \mathbf{x} + (1 - \mathbf{m}^i) \mathbf{p}^i$. Since DDNM leverages a binary mask by design and the pattern $\mathbf{p}^i$ is less relevant to the restoration, we resort to restore $\mathbf{m}^i \mathbf{x}$ instead of $\mathbf{x}_{\mathbf{m}^i, \mathbf{p}^i}$. \CR{More details of how DDNM restores $\mathbf{x}$ from its masked version are in the Appendix}. Examples of the image restoration are shown in Figure~\ref{fig:example}. After being processed by the diffusion model, the image restored from trojan can be fed into its further use as other clean images, e.g., testing the downstream classifier.

\section{Evaluation}
\label{sec:evaluation}

\begin{table}[!t]\setlength{\tabcolsep}{1.6pt}\renewcommand{\arraystretch}{1}
  \centering
 \fontsize{5}{8}\selectfont
  \caption{Performance of our {\name} in removing the trojan from test inputs.}
  \resizebox{\columnwidth}{!}{
  \begin{tabular}{|c|c|c|c|c|c|c|c|c|c|} \hline
        \multirow{2}{*}{\makecell{Pre-training\\dataset}} & \multirow{2}{*}{\makecell{Downstream\\dataset}} & \multicolumn{2}{c|}{\makecell{Clean encoders,\\without {\name}}} & \multicolumn{2}{c|}{\makecell{Clean encoders,\\with {\name}}} & \multicolumn{2}{c|}{\makecell{Trojaned encoders,\\without {\name}}} & \multicolumn{2}{c|}{\makecell{Trojaned encoders,\\wtih {\name}}} \\ \cline{3-10} 
        
        && \makecell{ACC (\%)} & \makecell{ASR (\%)} & \makecell{ACC (\%)} & \makecell{ASR (\%)} & \makecell{ACC (\%)} & \makecell{ASR (\%)} & \makecell{ACC (\%)} & \makecell{ASR (\%)}  \\ \hline \hline
        
        \multirow{3}{*}{CIFAR10} &
            \multirow{1}{*}{STL10} 
            & 74.38 & 1.01 & 74.00 & 2.25 & 73.75 & 100.0 & 73.15 & 1.43  \\ \cline{2-10}
            
            & \multirow{1}{*}{SVHN} 
            & 54.13 & 23.44 & 54.08 & 23.18 & 60.29 & 99.98 & 60.23 & 18.86 \\ \cline{2-10}
            
            & \multirow{1}{*}{EuroSAT} 
            & 74.41 & 3.52 & 74.37 & 1.67 & 75.59 & 100.0 & 75.44 & 1.13 \\ \hline
            
        \multirow{3}{*}{STL10} &
            \multirow{1}{*}{CIFAR10} 
            & 82.38 & 3.01 & 82.08 & 0.49 & 81.79 & 100.0 & 81.71 & 3.54  \\ \cline{2-10}
            
            & \multirow{1}{*}{SVHN} 
            & 46.37 & 23.95 & 46.11 & 27.54 & 50.10 & 100.0 & 49.95 & 25.66 \\ \cline{2-10}
            
            & \multirow{1}{*}{EuroSAT} 
            & 77.78 & 3.07 & 77.85 & 1.07 & 77.56 & 99.67 & 77.41 & 0.01 \\ \hline
            
  \end{tabular}}
\label{table:badencoder}
\end{table}

\subsection{Experiment Settings}
\label{sec:experiment_settings}

\myparatight{Dataset and models} We consider CIFAR10~\cite{krizhevsky2009learning} and STL10~\cite{coates2011analysis} as the datasets to pre-train self-supervised learning encoder. When CIFAR10 is used as the pre-training dataset, we use STL10, SVHN~\cite{Netzer2011svhn}, and EuroSAT~\cite{helber2019eurosat} as the downstream datasets. When STL10 is applied to pre-trained the encoder, we use CIFAR10, SVHN, and EuroSAT as the downstream dataset. We resize all images to be 32$\times$32 to be consistent. The details of these datasets can be found in Table 5 in the Appendix. Unless otherwise mentioned, we respectively use CIFAR10 and STL10 as the default pre-training and downstream dataset.  For details about pre-training encoders or training downstream classifiers, please refer to the Appendix. 

\myparatight{Attack settings}We consider attacks that poison the pre-training dataset~\cite{saha2022backdoor,liu2022poisonencoder} or manipulate the parameters of an encoder~\cite{jia2021badencoder}. As the attack that manipulates the encoder parameters is usually more powerful, we consider~\cite{jia2021badencoder} by default. We adopt the publicly available implementation of~\cite{saha2022backdoor,jia2021badencoder} and implement~\cite{liu2022poisonencoder} by ourselves. We use the default parameter settings in those works. In particular, we set the default trigger size  to 10$\times$10. Moreover, we use randomly generated trigger patterns.

\myparatight{Defense settings} Our {\name} has two parameters: $k$ (mask size) and $s$ (step size). By default, we set $k$ and $s$ to 15 and 1, respectively. We will study the impact of each of them in the ablation study. For image restoration, we use a pre-trained diffusion model from~\cite{ddnm-url}. 

\myparatight{Evaluation metrics} To evaluate the effectiveness of our approach in detecting trojaned test images, we use the \textit{False Positive Rate} (FPR) and \textit{False Negative Rate} (FNR) as metrics. FPR (or FNR) measures the fraction of clean (or trojan) test images that are detected as trojan (or clean). Additionally, to measure the effectiveness of {\name} in recovering trojaned images, we apply our method to both the training and testing data of the downstream classifier and use the \textit{Accuracy} (ACC) and \textit{Attack Success Rate} (ASR) to measure the classifier's accuracy and attack success. Specifically, ACC measures the prediction accuracy of the downstream classifier for clean test images, while ASR quantifies the fraction of trojaned images predicted as the target class.

\begin{table}[!t]\setlength{\tabcolsep}{2pt}\renewcommand{\arraystretch}{1}
  \centering
  \fontsize{5}{8}\selectfont
  \caption{Performance of {\name} in detecting trojaned test inputs. FNR is undefined for clean encoders as there is no trojaned example when the encoder is clean. }
        \begin{tabular}{|c|c|c|c|c|c|} \hline
        
        \multirow{2}{*}{\makecell{Pre-training\\dataset}} & \multirow{2}{*}{\makecell{Downstream\\dataset}} & \multicolumn{2}{c|}{\makecell{Clean encoders}} & \multicolumn{2}{c|}{\makecell{Trojaned encoders}} \\ \cline{3-6}
        
        && \makecell{FPR (\%)} & \makecell{FNR (\%)} & \makecell{FPR (\%)} & \makecell{FNR (\%)} \\ \hline \hline
        
        \multirow{3}{*}{CIFAR10} &
            \multirow{1}{*}{STL10} 
            & 0.88 & - & 1.02 & 0.40     \\ \cline{2-6}
            
            & \multirow{1}{*}{SVHN} 
            & 0.15 & - & 0.11 & 0.00     \\ \cline{2-6}
            
            & \multirow{1}{*}{EuroSAT} 
            & 0.04 & - & 0.29 & 1.12     \\ \hline
        
        \multirow{3}{*}{STL10} &
            \multirow{1}{*}{CIFAR10} 
            & 0.80 & - & 0.41 & 0.00     \\ \cline{2-6}
            
            & \multirow{1}{*}{SVHN} 
            & 1.11 & - & 0.49 & 0.01     \\ \cline{2-6}
            
            & \multirow{1}{*}{EuroSAT} 
            & 0.07 & - & 0.26 & 0.00     \\ \hline
            
        \end{tabular}
\label{table:detection}
\end{table}

\subsection{Defending against State-of-the-art Attacks}
\label{sec:experiment_main_results}

\myparatight{Parameter manipulation attacks}
Table~\ref{table:detection} and~\ref{table:badencoder} present the performance of our {\name} in trojan detection and removal against trojan attack that manipulate the parameters of encoders~\cite{jia2021badencoder}. We have three observations from the results. First, our {\name} effectively detects trojaned test inputs, as evidenced by the low FNR in Table~\ref{table:detection}. Second, {\name} can successfully remove trojans from trojaned inputs. In Table~\ref{table:badencoder}, the ASR values significantly reduce after applying our method to trojaned encoders, compared to trojaned encoders without {\name}. For instance, for the trojaned encoder pre-trained on CIFAR10 and the downstream dataset being STL10, the ASR without {\name} is 100\%, whereas the ASR with {\name} is decreased to 1.43\%. It is important to note that while the ASR with {\name} are not zero, they are comparable to the ASR of clean encoders. This is because downstream classifiers can misclassify some test inputs, leading to slightly higher ASR. Our third observation is that {\name} successfully maintains the utility of the encoders. Regardless of whether the encoder is clean or trojaned, the FPR values remain small and the ACC values are comparable to those without {\name}. Overall, our {\name} shows promising performance in detecting and removing trojans from test images (if any), while maintaining the utility of the encoders.

\myparatight{Pre-training dataset poisoning attacks} We also evaluate our method on trojan attacks that poison the pre-training dataset~\cite{saha2022backdoor,liu2022poisonencoder}. For each attack, we craft a trojaned encoder pre-trained on CIFAR10 and set the target downstream task to be STL10. The results in Table 11a and 11b in the Appendix show our {\name} can defend against these trojan attacks effectively.

\begin{figure}[!t]
	 \centering
\subfloat[Beatrix]{\includegraphics[width=0.18\textwidth]{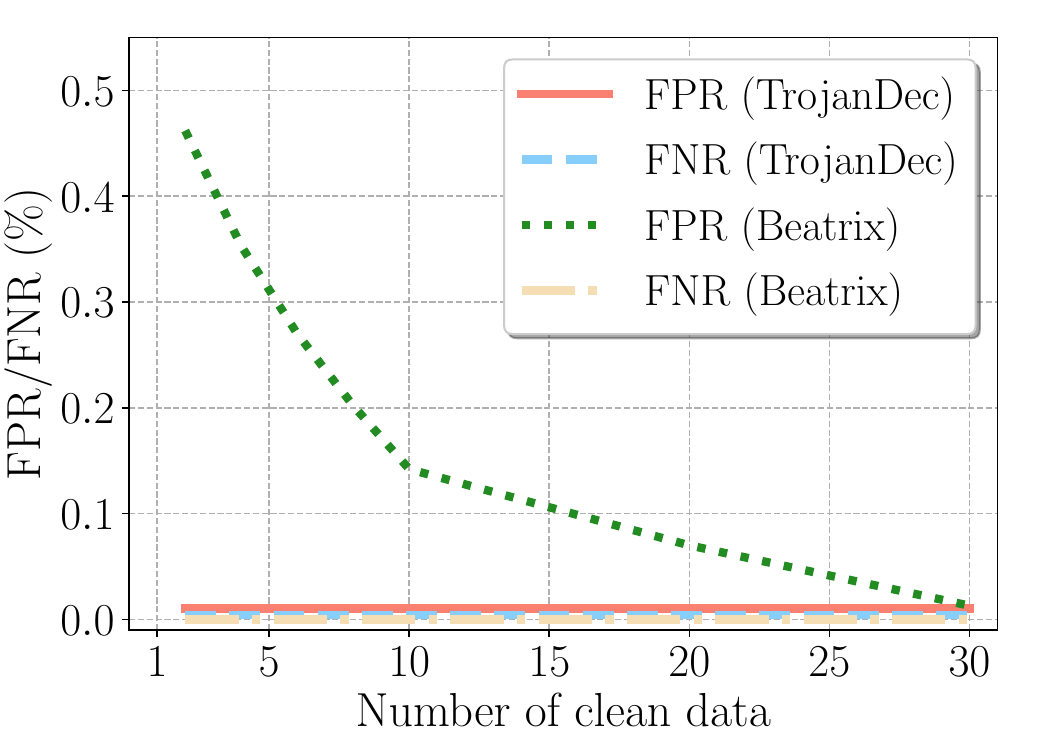}\label{fig:beatrix_stl10}}
\subfloat[Strip]{\includegraphics[width=0.18\textwidth]{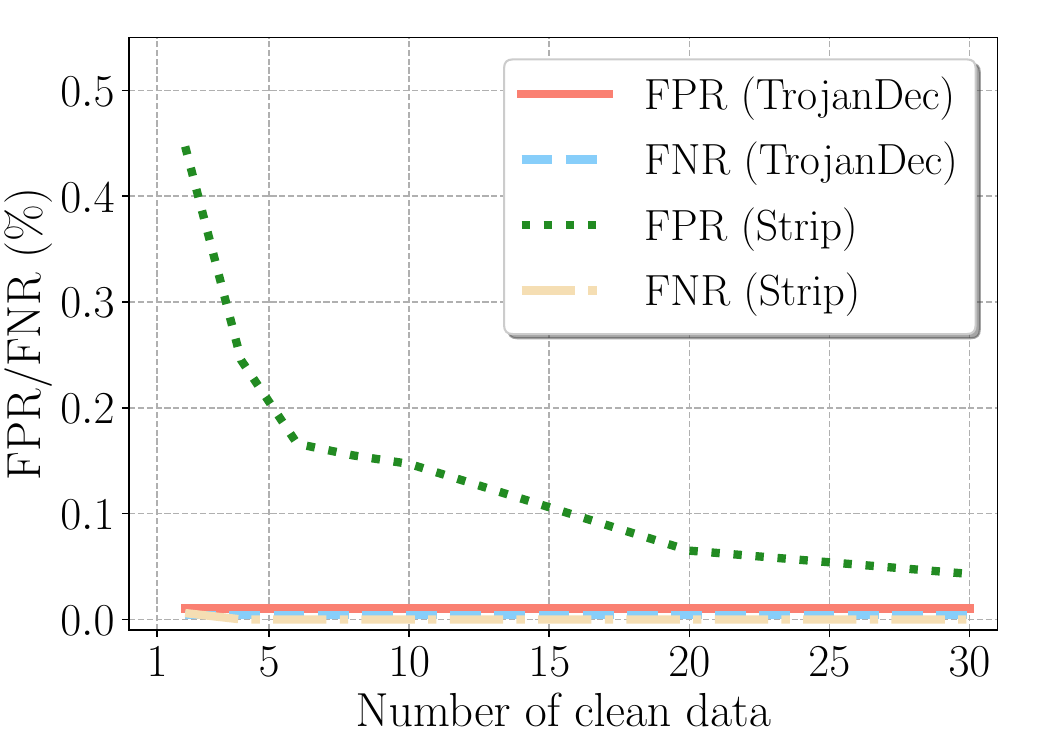}\label{fig:strip_stl10}}
\caption{Comparison to Beatrix and Strip on STL10. }
\label{fig:beatrix_and_strip_stl10}
\end{figure}

\subsection{Comparing {\name} with Existing Defenses} We compare our {\name} with several state-of-the-art testing-phase defenses~\cite{gao2019strip,doan2020februus,ma2022beatrix}. In addition, we show existing state-of-the-art training-phase defenses~\cite{wang2019neural,xu2019detecting,zeng2022ibau} are not effective. We leverage their publicly available code in our experiments.  

\myparatight{Beatrix~\cite{ma2022beatrix}}
Beatrix utilizes Gramian information to distinguish trojaned inputs from clean ones. Beatrix requires the defender to have enough clean validation images. However, Beatrix is not effective when the defender does not have enough clean validation images. To compare with Beatrix, we assume that the defender has varying numbers of clean validation images and plot the results on STL10 in Figure~\ref{fig:beatrix_stl10}. As {\name} does not rely on clean validation images, the FPR and FNR remain unchanged irrespective of the number of clean validation images. We have two observations from the results. First, our {\name} achieves comparable performance with Beatrix when the defender has enough clean validation images (e.g., more than 30 clean validation images). Second, our {\name} significantly outperforms Beatrix when the clean validation images are insufficient. 
We have similar observations on other downstream datasets, as shown in Figures 5a and 6a in the Appendix. 

\myparatight{Strip~\cite{gao2019strip}} Strip also leverages clean validation images to detect whether a test input is trojaned or not. Figure~\ref{fig:strip_stl10} compares Strip's performance on STL10 with that of {\name}. Similar to Beatrix, when the defender has limited clean validation images, Strip's FPR is high. As clean validation images number increases, Strip's performance improves. We find that our {\name} still outperforms Strip even if we assume the defender has 30 clean validation images.
Similar trends were observed on other downstream datasets, as presented in Figure 5b and 6b in the Appendix. 

\myparatight{Februus} Februus~\cite{doan2020februus} is a defense method that uses GradCAM~\cite{selvaraju2017gradcam} to determine the potential trigger area. Then, it applies a Wasserstein GAN~\cite{gulrajani2017wasserstein} for in-painting the detected area. Table 6 in the Appendix shows the comparison results. Our {\name} achieves slightly better ACC and lower ASR for the trojaned inputs. Besides, we note that Februus has two limitations compared to our method. Firstly, it cannot detect if a test input contains the trojan trigger, but can only directly restore every test input. This makes their defense method limited in scenarios where the trojan behaviours are expected to be reported. Secondly, its restoration technique requires the defender's knowledge of the distribution of the real unmask testing data, which is not available in the data-free setting we consider.

\myparatight{Other defenses}
We also evaluate the state-of-the-art training-phase defense methods~\cite{wang2019neural,xu2019detecting,zeng2022ibau} against trojan attacks. Table 7 in the Appendix shows these methods fail to defend against trojan attacks to encoders. Our experimental results confirm the observations in~\cite{carlini2021poisoning,jia2021badencoder,feng2023detect}.

\subsection{Real-world Encoders}
\label{sec:experiment_clip}

To further evaluate the effectiveness of {\name}, we apply it on 2 real-world self-supervised learning encoders: 1) the encoder pre-trained on ImageNet released by Google~\cite{simclr-url} and 2) CLIP encoder pre-trained on 400 million image-text pairs released by OpenAI~\cite{clip-url}. Both encoders take 224$\times$224 images as inputs. To fit its input size, we set $k$ to 150 and $s$ to 10. We directly apply our method on trojaned CLIP encoders released in~\cite{badencoder-url}. The trojan detection and removal results presented in Table~\ref{table:case_study} show that {\name} is effective in defending against trojan attacks to large, real-world pre-trained self-supervised learning encoders. More experimental results on larger-scale datasets are in Table 8 in the Appendix.

\begin{table}[!t]\setlength{\tabcolsep}{2pt}\renewcommand{\arraystretch}{1}
  \centering
  \fontsize{5}{8}\selectfont
  \caption{Performance of {\name} in defending against trojan attacks to 2 real-world encoders.}
  \subfloat[Trojan detection]{
        \begin{tabular}{|c|c|c|c|} \hline
        
        \makecell{Pre-training\\dataset} & \makecell{Downstream\\dataset} & \makecell{FPR (\%)} & \makecell{FNR (\%)} \\ \hline \hline
        
        \multirow{3}{*}{ImageNet} 
            & STL10 & 1.16 & 0.00     \\ \cline{2-4}
            & SVHN & 0.10 & 0.20     \\ \cline{2-4}
            & GTSRB & 3.20 & 0.86     \\ \cline{1-4}
            
        \multirow{3}{*}{CLIP} 
            & STL10 & 1.65 & 1.16     \\ \cline{2-4}
            & SVHN & 0.94 & 0.02     \\ \cline{2-4}
            & GTSRB & 1.34 & 1.04     \\ \cline{1-4}
            
        \end{tabular}
  \label{table:case_study_detection}
  }
  
  \subfloat[Trojan removal]{
        \begin{tabular}{|c|c|c|c|c|c|} \hline
        \multirow{2}{*}{\makecell{Pre-training\\dataset}} & \multirow{2}{*}{\makecell{Downstream\\dataset}} & \multicolumn{2}{c|}{\makecell{Attacked encoders\\without {\name}}} & \multicolumn{2}{c|}{\makecell{Attacked encoders,\\wtih {\name}}} \\ \cline{3-6} 
        
        && \makecell{ACC (\%)} & \makecell{ASR (\%)} & \makecell{ACC (\%)} & \makecell{ASR (\%)}  \\ \hline \hline
        
        \multirow{3}{*}{ImageNet} 
            & STL10 & 95.84 & 99.99 & 95.40 & 4.28  \\ \cline{2-6}
            & SVHN & 68.06 & 100.0 & 68.04 & 8.12  \\ \cline{2-6}
            & GTSRB & 47.41 & 99.83 & 45.84 & 9.16  \\ \cline{1-6}
            
        \multirow{3}{*}{CLIP} 
            & STL10 & 96.26 & 99.96 & 95.36 & 1.94  \\ \cline{2-6}
            & SVHN & 61.54 & 100.0 & 61.16 & 0.06  \\ \cline{2-6}
            & GTSRB & 59.58 & 97.02 & 59.30 & 3.82 \\ \hline
            
        \end{tabular}
    \label{table:case_study_removal}
    }
\label{table:case_study}
\end{table}

\subsection{Adaptive Attacks}
\label{sec:experiment_adaptive}

In this section, we consider several adaptive attacks the adversary can take to evade our defense. In summary, we find that {\name} can effectively defend against adaptive attacks.

\myparatight{Trigger with large size}
The first countermeasure we consider is that the attacker uses a large trigger such that the trigger size is bigger than the mask size set by {\name}. Table 9 in the Appendix presents the defense results where the trigger size is 16$\times$16. The FNR values have slightly increased compared to Table~\ref{table:detection} where the trigger size is 10$\times$10. Also, the ASR after applying {\name} is decreased to small values. This is because our method does not necessarily require the mask to fully cover the trigger. As shown in Figure~\ref{dist}, only one fourth of the trigger being covered by the mask is enough to divide the metadata into two clusters. Our results demonstrate that our {\name} can effective defend against trojan attacks with large trigger sizes.

\myparatight{Trigger with dynamic location}
We also consider an attacker that performs a dynamic trojan attack to the encoder. Since there is no existing work studying dynamic trojan attacks to encoders, we generalize the idea from~\cite{salem2020dynamic} to perform the dynamic trojan attack. In particular, we inject the trojan into the encoder in the same way as~\cite{jia2021badencoder} and in the testing phase, we embed the trigger into the test input at a random location. We evaluate the effectiveness of our {\name} against these two adpative attacks on 3 encoders pre-trained on CIFAR10 and use STL10 as the downstream task. Table~\ref{table:dynamic} presents the  results, which shows that our {\name} can effectively defend against trojan attacks with dynamic trigger location.

\myparatight{Trojan attacks in supervised learning~\cite{chen2017targeted,yao2019latent,ning2021cleanlabel}}
We notice that in supervised learning, there exist other types of trojan attacks, e.g., blending attack~\cite{chen2017targeted}, latent attack~\cite{yao2019latent}, and clean-label attack~\cite{ning2021cleanlabel}. For the blending attack, we generalize it into the self-supervised learning domain by using the default trigger presented in~\cite{chen2017targeted}. However, as Table 10 in the Appendix reveals, the attack is not very successful, as the accuracy after the trojan injection decreases too much, and the ASR is not high for STL10 and EuroSAT. This indicates that it is not trivial to generalize the existing blending trojan attacks to the self-supervised learning context. 
For latent and clean-label trojan attacks, since they require label information to perform the attack, they are not applicable in attacking encoders pre-trained on unlabeled data.

\subsection{Ablation Study}
\label{sec:experiment_ablation}

\myparatight{Impact of $k$ and $s$}
Our {\name} has two parameters: $k$ and $s$, which control the mask size and the step size used to generate the mask set in the metadata extraction component. Figure 7 and 8 in the Appendix show the impact of $k$ and $s$ respectively. When the value of $k$ is small, e.g., less than 17, both the trojan detection and removal achieve a consistently good performance. When $k$ is large, we find that FNR and ASR increase. The reason for this trend is that when $k$ is large, it is likely that almost all masks in the mask set will cover a significant portion of the real trigger, resulting in none of the masked variants containing a complete trigger. This leads to all masked variants being not similar to the original trojaned input, making the detection component less effective.
We have a similar observation for $s$. Our results demonstrate that we can set a smaller $k$ and $s$ in practice.

\begin{table}[!t]\setlength{\tabcolsep}{2pt}\renewcommand{\arraystretch}{1}
  \centering
  \fontsize{5}{8}\selectfont
  \caption{Performance of {\name} against adapted dynamic trojan attacks.} 
  \subfloat[Trojan detection]{
        \begin{tabular}{|c|c|c|} \hline
        
        \makecell{Downstream\\dataset} & \makecell{FPR (\%)} & \makecell{FNR (\%)} \\ \hline \hline
        
        STL10 & 0.70 & 0.65     \\ \cline{1-3}
        SVHN & 0.57 & 1.20     \\ \cline{1-3}
        EuroSAT & 2.35 & 3.12     \\ \cline{1-3}
            
        \end{tabular}
  \label{table:detecting_dynamic}
  }
  
  \subfloat[Trojan removal]{
        \begin{tabular}{|c|c|c|c|c|} \hline
        \multirow{2}{*}{\makecell{Downstream\\dataset}} & \multicolumn{2}{c|}{\makecell{Trojaned encoders\\without {\name}}} & \multicolumn{2}{c|}{\makecell{Trojaned encoders,\\wtih {\name}}} \\ \cline{2-5} 
        
        & \makecell{ACC (\%)} & \makecell{ASR (\%)} & \makecell{ACC (\%)} & \makecell{ASR (\%)}  \\ \hline \hline
        
        \multirow{1}{*}{STL10} 
            & 73.75 & 99.98 & 73.45 & 1.64  \\ \cline{1-5}
            
        \multirow{1}{*}{SVHN} 
            & 60.29 & 98.41 & 59.76 & 23.75 \\ \hline
            
        \multirow{1}{*}{EuroSAT} 
            & 75.59 & 99.07 & 73.85 & 20.96 \\ \hline
            
        \end{tabular}
    \label{table:removing_dynamic}
    }
\label{table:dynamic}
\end{table}
\section{Conclusion and Future Work}
\label{sec:conclusion}

In this work, we propose {\name}, a data-free framework to defend against trojan attacks to self-supervised learning. Specifically, the defender first extracts metadata from a given test input and identifies whether it has malicious trigger using statistical analysis. Then, the defender can restore the trojaned test input via a pre-trained diffusion model. Also, our extensive evaluation results show that {\name} is effective against multiple types of trojan attacks, and outperforms the existing trojan defense mechanisms, especially in the setting where the defender only has access to a limited number of clean data. Moreover, we demonstrate that several countermeasures fail to break {\name}. Interesting future works would be to extend our framework to defend against trojan attacks \CR{with non-patch-based triggers or other domains}.

\section{Acknowledgments}

\CR{We thank the anonymous reviewers for insightful reviews.}

\bibliography{ref}
\appendix

\section{\CR{Discussion}}

\subsection{Trojan Attacks}

\myparatight{Trojan attacks in supervised learning} In addition to the trojan attacks evaluated in our experiments, there are many recent works studying trojan attacks to supervised learning models~\cite{ning2021cleanlabel,jha2023labelpoisoningneed,wu2023efficient,boberirizar2023architectural,jiang2023color,yuan2023catching}. Since they focus on supervised learning and rely on label information to implant the trojan, they are not applicable to self-supervised learning settings in general. 

\myparatight{Trojan attacks in self-supervised learning} Existing trojan attacks to self-supervised learning encoders can be categorized into two types: 1) poisoning the pre-training dataset~\cite{saha2022backdoor,liu2022poisonencoder,carlini2021poisoning,li2023embarrassingly,li2024on,zhang2022corruptencoder,sun2024backdoor,bai2023badclip}, and 2) directly manipulating the encoder parameters~\cite{jia2021badencoder}. In this work, we have evaluated both types of attacks and showed that our \name{} achieve a good performance in detecting and restoring them. In this section, we further evaluate CorruptEncoder~\cite{zhang2022corruptencoder}, which is a more advanced attack technique that is able to gain a high trojan effectiveness by only poisoning a small fraction (i.e., $\leq$ 5\%) of pre-training dataset. In particular, we directly apply our \name{} under default settings to the poisoned encoder they released, which was pre-trained on a subset of ImageNet with 100 classes and the downstream dataset is another subset of ImageNet with another 100 classes. The FPR and FNR are 3.78 and 1.33, which means that our \name{} is effective in detecting trojan inputs crafted by CorruptedEncoder. The ASR before and after applying our \name{} are 99\% and 3\%, respectively, which shows that our \name{} can successfully restore the trojan inputs. 

\subsection{Defenses against Trojan Attacks}

\myparatight{Defenses against trojan attacks in self-supervised learning} There are many existing studies about training-phase defenses against trojan attacks~\cite{chen2018detecting,wang2019neural,kolouri2020litmus,huang2022decoupling,zheng2022lipschitzness,tejankar2023defending,feng2023detect,liu2023beating,zheng2024sslcleanse,wang2024mmdb,hritik2023cleanclip,du2024defendingdeepregressionmodels}. As a testing-phase defense approach, our \name{} can complement these training-phase approaches. For instance, CLP is a training-time pruning method which leverages the Lipschitz constant to remove the trojan channels from the models~\cite{zheng2022lipschitzness}. SSL-Cleanse is another training-time defense aimed at detecting and mitigating Trojaned models~\cite{zheng2024sslcleanse}, while our method operates at run-time, detecting if a specific test example activates the Trojan behavior of a model and then recovering the test example itself rather than the model. 

\myparatight{Other defenses} There exist other defenses against trojan attacks. For instance, BEAGLE focuses on the forensics of deep learning backdoor attacks from a perspective different from ours~\cite{cheng2023beagle}. MDP is a testing-phase trojan input detection technique in the language domain~\cite{xi2024mdp}, while our method is centered in the vision domain. Django focuses on detecting trojan inputs in object detection tasks~\cite{shen2023django}. BIRD~\cite{chen2023bird} and PolicyCleanse~\cite{guo2023policy} concentrate on defending against trojans in the reinforcement learning domain. 

\section{Proof of Proposition~\ref{prop_1}}
\label{proof_of_proposition_1}

Given a mask $(\mathbf{m}, \mathbf{p})$ where $\mathbf{p}$ is randomly generated, we use $p'$ to denote the probability that there exists a part of the mask such that $\ell_1$ distance of this part of the mask pattern and the trojan trigger is no larger than $\beta$. As we randomly sample each entry of $\mathbf{p}$ based on a uniform distribution between $[0,1]$, $p'$ is no larger than the volume of an $\ell_1$-ball with radius $\beta$ in the space $\mathbb{R}^T$, where $T = e_h \cdot e_w$. In other words, we have $p' \leq \frac{(2\beta)^T}{T!}$. The probability that there exists a part of the mask such that $\ell_1$ distance of this part of the mask pattern and the trojan trigger is larger than $\beta$ can be computed as $1-p'$. Based on the fact that $p' \leq \frac{(2\beta)^T}{T!}$, we know this probability is no smaller than $1 - \frac{(2\beta)^T}{T!}$. We reach the conclusion.

\section{Details of DDNM}
\label{ddnm_downsampling}
The goal of the diffusion model in general is to train a model to estimate the noise added to $\mathbf{x}_0$ based on $\mathbf{x}_t$. To reach the goal, it uses the following loss function:
\begin{align}
\label{eq:diffusion_loss}
  L_\theta = \lVert \epsilon - \epsilon_\theta(\sqrt{\bar{\alpha}_t}\mathbf{x}_0 + \epsilon\sqrt{1-\bar{\alpha}_t}, t) \rVert^2_2,
\end{align}
where $\epsilon_\theta$ is the neural backbone, $t$ represents a time step, $\bar{\alpha}_t$ is pre-defined scalar constant, and $\epsilon \sim \mathcal{N}(0, \mathbf{I})$ represents the zero-mean Gaussian noise. 

For simplicity, we denote $\mathbf{A}=\mathbf{m}^i$. In the image restoration task, we have $\mathbf{y} = \mathbf{A}\mathbf{x}$, where $\mathbf{y}$ is the degraded image and $\mathbf{A}$ is a linear degradation operation (i.e., masking). Given the degraded image $\mathbf{y}$, our goal switches to output $\hat{\mathbf{x}}$ which is the estimation of $\mathbf{x}$. From $\mathbf{A}\hat{\mathbf{x}} = \mathbf{y}$, we obtain that $\hat{\mathbf{x}} = \mathbf{A}'\mathbf{y} + (\mathbf{I} - \mathbf{A}'\mathbf{A})\bar{\mathbf{x}}$, where $\mathbf{A}'$ is the inverse of $\mathbf{A}$. Moreover, since $\mathbf{y}$ is known, our goal is to generate $\bar{\mathbf{x}}$ such that the null-space part $(\mathbf{I} - \mathbf{A}'\mathbf{A})\bar{\mathbf{x}}$ is in the same distribution as the range-space part $\mathbf{A}'\mathbf{y}$.  To achieve this goal, DDNM changes the way that $\mathbf{x}_{t-1}$ is sampled. Instead of sampling $\mathbf{x}_{t-1}$ from $p(\mathbf{x}_{t-1}|\mathbf{x}_t, \mathbf{x}_0)$ which yields noisy intermediate states, DDNM first outputs an estimation to $\mathbf{x}_0$ at time step $t$:

\begin{align}
\label{eq:est_x0_t}
  \mathbf{x}_{0|t} = \frac{1}{\sqrt{\bar{\alpha}}_t}(\mathbf{x}_t - \epsilon_\theta(\mathbf{x}_t, t)\sqrt{1 - \bar{\alpha}_t})
\end{align}

Then, in order to finally output a $\mathbf{x}_0$ that satisfies $\mathbf{A}\mathbf{x}_0 = \mathbf{y}$, DDNM fixes the range-space as $\mathbf{A}'\mathbf{y}$ and yields a refined estimation $\hat{\mathbf{x}}_{0|t} = \mathbf{A}'\mathbf{y} + (\mathbf{I} - \mathbf{A}'\mathbf{A})\mathbf{x}_{0|t}$. Thus, instead of sampling from $p(\mathbf{x}_{t-1}|\mathbf{t}_t, \mathbf{x}_0)$, DDNM samples $\mathbf{x}_{t-1}$ from $p(\mathbf{x}_{t-1}|\mathbf{x}_t, \hat{\mathbf{x}}_{0|t})$:

\begin{align}
\label{eq:ddnm_sample}
  \mathbf{x}_{t-1} = \frac{\sqrt{\bar{\alpha}_{t-1}}\beta_t}{1 - \bar{\alpha}_t}\hat{\mathbf{x}}_{0|t} + \frac{\sqrt{\alpha_t}(1 - \bar{\alpha}_{t-1})}{1 - \bar{\alpha}_{t}}\mathbf{x}_t + \sigma_t\epsilon
\end{align}

By applying Equations~\ref{eq:est_x0_t} and~\ref{eq:ddnm_sample} iteratively, DDNM is pre-trained on the loss function~\ref{eq:diffusion_loss}.

\section{Pre-training Encoders and Training Downstream Classifiers}
\label{pretraining_details}
For the self-supervised learning pre-training, we use the public code of SimCLR~\cite{simclr-url} and use its default setting. In particular, we use the ResNet-18~\cite{he2016deep} as the backbone of the self-supervised learning encoders. The algorithm we use to pre-train the encoders is SimCLR~\cite{chen2020simple}. The pre-training epoch number is 1,000 and the batch size is 256. We use the Adam optimizer with the initial learning rate being 0.001. The data augmentation operations used to process the pre-training data include RandomResizedCrop,  RandomHorizontalFlip,  ColorJitter, and RandomGrayScale. During the pre-training phase with CIFAR10 (or STL10), we use its training data to train the encoders. 

For the downstream evaluation, following previous works~\cite{he2016deep,liu2022poisonencoder}, we use the linear classifier as the downstream classifier. Specifically, we use the Adam optimizer with an initial learning rate 0.0001 and batch size 256 to train a downstream classifier for 100 epochs, with no data augmentation operation being used. For a downstream dataset (e.g., SVHN), we use its train set for training the downstream classifier and use the test set for evaluating ACC (or ASR). 

\section{Cost of our \name{}}
\label{cost_analysis}
\CR{Following prior work~\cite{li2023embarrassingly,jia2021badencoder,saha2022backdoor}, we consider that an attacker releases a trojaned encoder on open-source platforms (e.g., Hugging Face) for users to download. Thus, our method incurs no additional financial cost for queries. High-performance open-source encoders, such as the CLIP encoder used in our experiments, are freely accessible. In terms of query time, we can make predictions for a batch of masked examples to reduce overhead. Under default settings, our method requires 1.9 ms to detect a trojaned test input (using an A100 GPU), which is comparable to the around 1 ms required by STRIP for a similar detection. }

\section{Necessity of Image Restoration}
\label{nec_image_restoration}
\CR{To demonstrate the necessity of the diffusion model and the image restoration step in our \name{}, we compared the accuracy of test images before and after recovery. In particular, we use the encoders pre-trained on CIFAR10 and use STL10, SVHN, and EuroSAT as downstream datasets. For STL10 / SVHN / EuroSAT, the accuracy before restoration is 65.71\% / 37.84\% / 70.29\%, while the accuracy after restoration is 73.15\% / 60.23\% / 75.44\%. As a result, the image restoration step improves the accuracy by 7\% / 22\% / 5\%. These accuracy improvements across various datasets highlight the diffusion model's effectiveness in enhancing detection accuracy and demonstrate the necessity of the recovery step. We will add results to our paper. }

\begin{algorithm}[tb]
   \caption{Mask Set Generation}
   \label{alg:mask_set}
\begin{algorithmic}[1]
   \STATE {\bfseries Input:} $k$ (mask size), $s$ (step size), and $t$ (image size).
   \STATE {\bfseries Output:}  $\mathcal{M}$ (a set of masks) \\
    
   \STATE $\mathcal{M} \gets \emptyset$ \\
   \STATE $a \gets 0$ \\

   \WHILE{$a+k \leq t$}
   \STATE $b \gets 0$ \\
   \WHILE{$b+k \leq t$} 
   
   \STATE $(\mathbf{m}, \mathbf{p}) \gets \textit{CreateMask}(a, b, k)$ \\
   \STATE $\mathcal{M}$.$\textrm{append}((\mathbf{m}, \mathbf{p}))$ \\
   \STATE $b=b+s$ \\
   
   \ENDWHILE
   
   \STATE $a=a+s$ \\
   \ENDWHILE
   
   \STATE \textbf{return} $\mathcal{M}$
\end{algorithmic}
\end{algorithm}

\begin{table}[tp]\renewcommand{\arraystretch}{1.2} 
\fontsize{7}{9}\selectfont
\centering
	\caption{Dataset summary.}
	\setlength{\tabcolsep}{1mm}
	{
	\begin{tabular}{|c|c|c|c|}
		\hline
	 \makecell{Dataset} & \makecell{Task} & \makecell{\#Training \\Examples} & \makecell{\#Testing \\Examples} \\ \hline \hline
	CIFAR10 & Object classification & 50,000 & 10,000  \\ \cline{1-4}  
	STL10 & Object classification & 5,000 & 8,000  \\ \cline{1-4} 
	SVHN & Street number classification & 73,257 & 26,032  \\ \cline{1-4}
	EuroSAT & Landscape classification & 24,300 & 2,700  \\ \cline{1-4}
	\end{tabular}
	}
	\label{dataset_table}
\end{table}

\begin{figure}[!t]
	 \centering
\subfloat[Beatrix]{\includegraphics[width=0.23\textwidth]{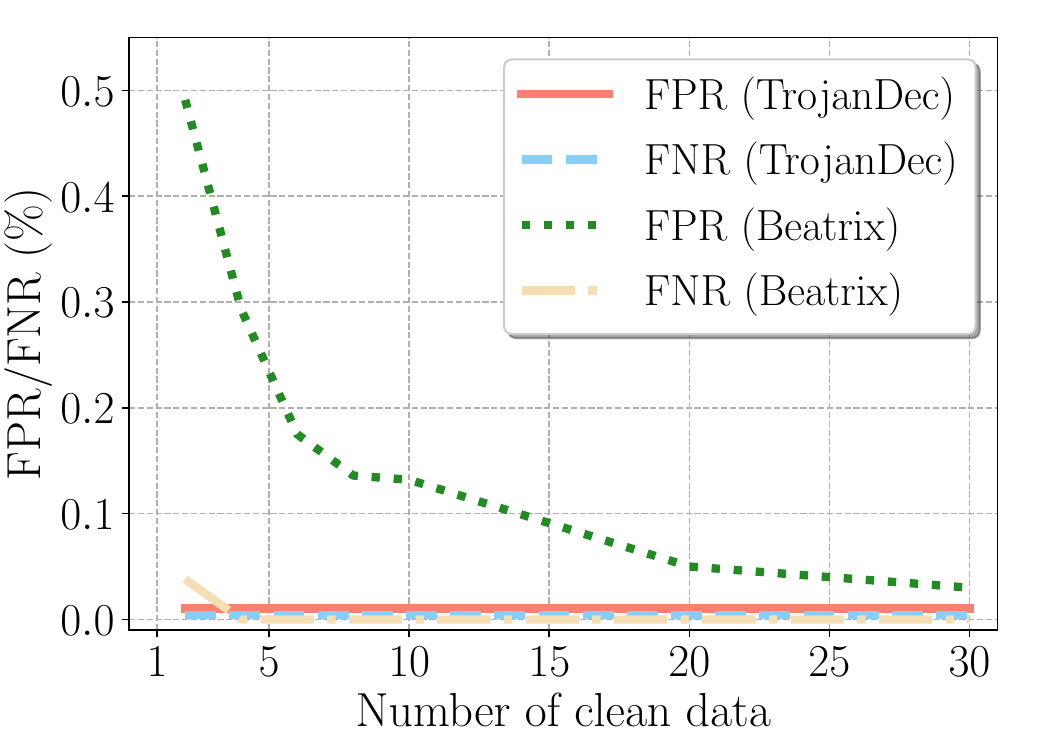}\label{fig:beatrix_svhn}}
\subfloat[Strip]{\includegraphics[width=0.23\textwidth]{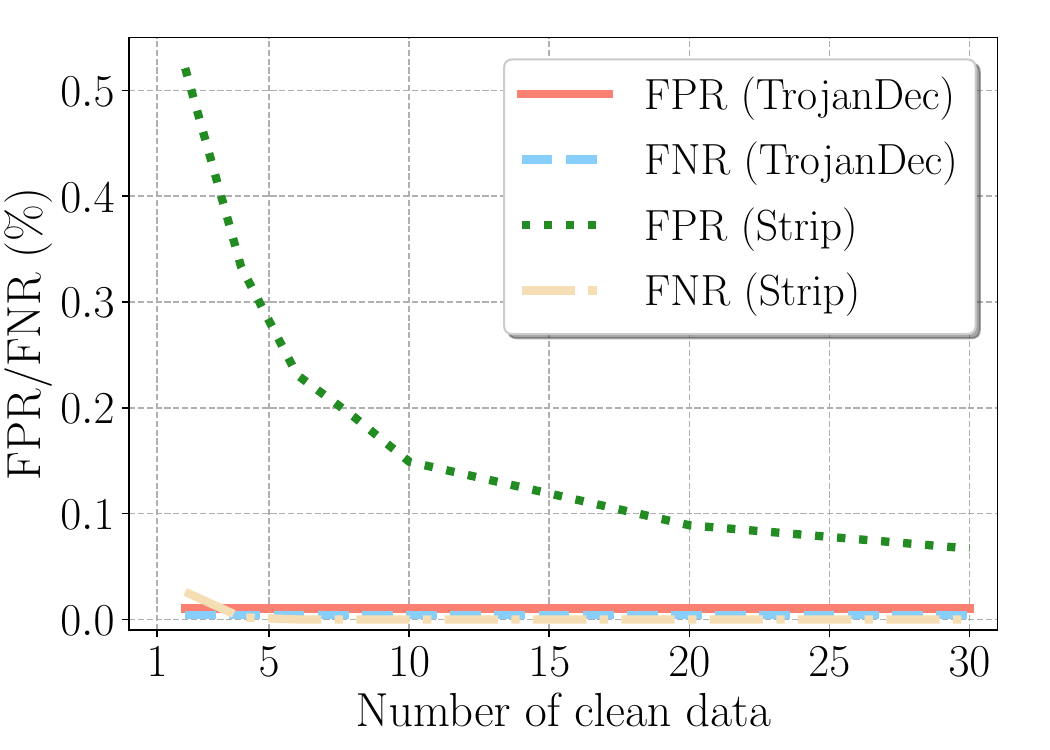}\label{fig:strip_svhn}}
\caption{Comparison to Beatrix and Strip on SVHN. }
\label{fig:beatrix_and_strip_svhn}
\end{figure}

\begin{figure}[!t]
	 \centering
\subfloat[Beatrix]{\includegraphics[width=0.23\textwidth]{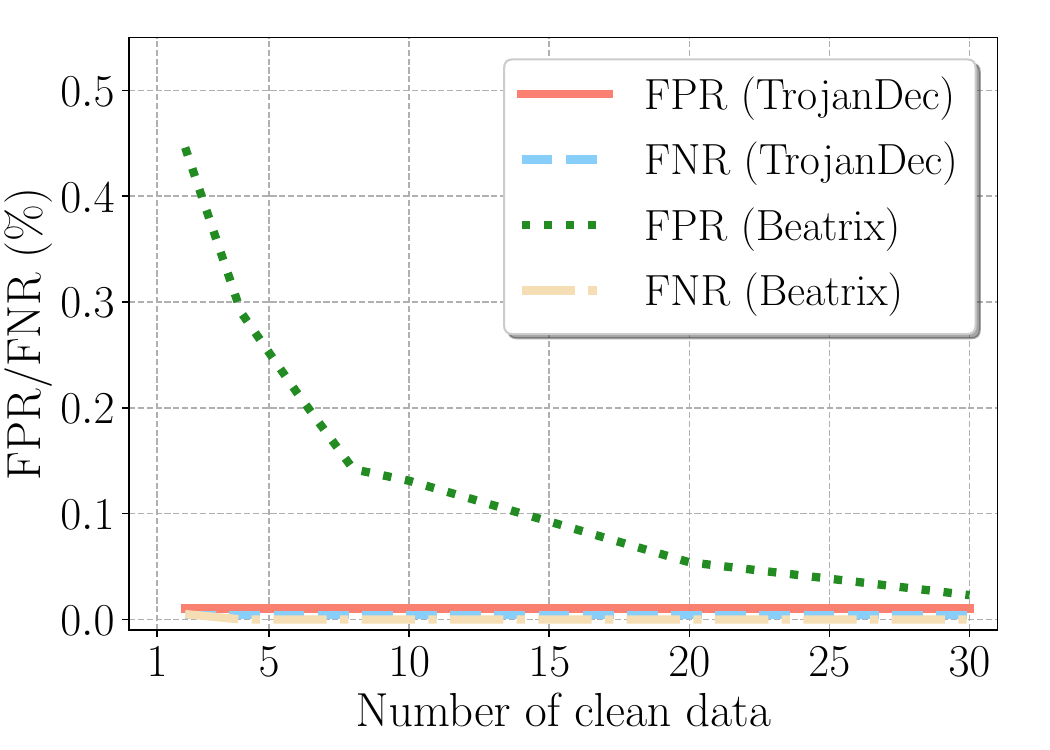}\label{fig:beatrix_eurosat}}
\subfloat[Strip]{\includegraphics[width=0.23\textwidth]{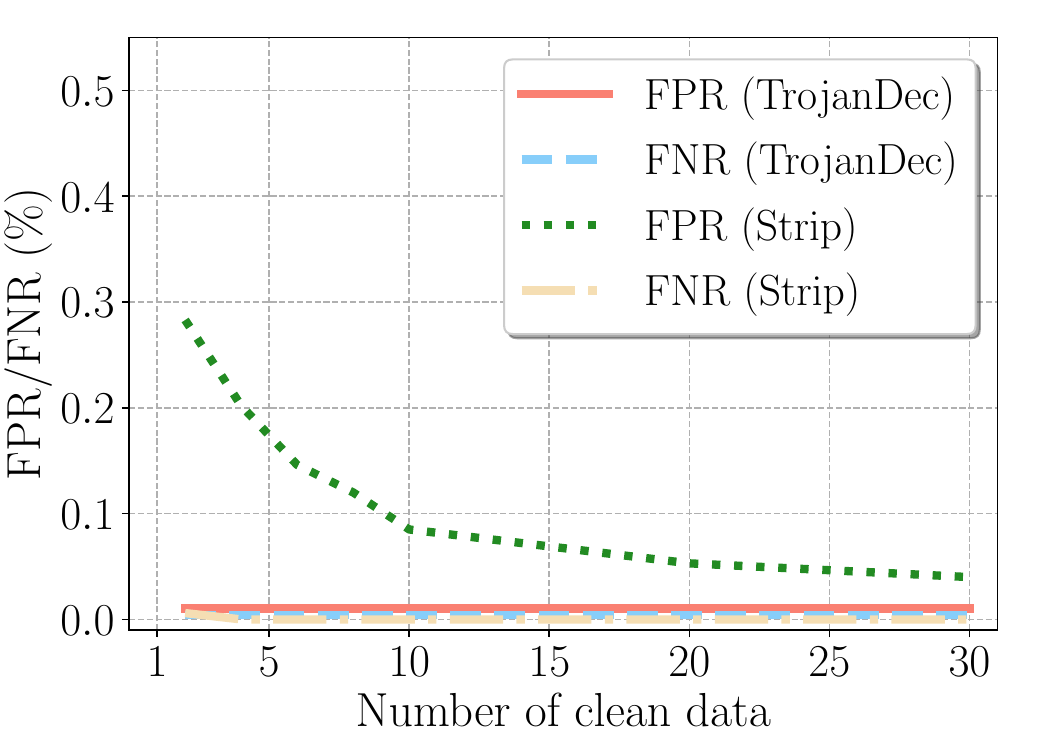}\label{fig:strip_eurosat}}
\caption{Comparison to Beatrix and Strip on EuroSAT. }
\label{fig:beatrix_and_strip_eurosat}
\end{figure}

\begin{table}[tp]\renewcommand{\arraystretch}{1.2} 
\fontsize{7}{9}\selectfont
\centering
	\caption{Comparison to Februus. }
	\setlength{\tabcolsep}{1mm}
	{
        \begin{tabular}{|c|c|c|c|c|} \hline
        \multirow{2}{*}{\makecell{Downstream\\dataset}} & \multicolumn{2}{c|}{\makecell{Trojaned encoders\\with Februus}} & \multicolumn{2}{c|}{\makecell{Trojaned encoders,\\wtih {\name}}} \\ \cline{2-5} 
        
        & \makecell{ACC (\%)} & \makecell{ASR (\%)} & \makecell{ACC (\%)} & \makecell{ASR (\%)}  \\ \hline \hline
        
        \multirow{1}{*}{STL10} 
            & 73.06 & 3.12 & 73.15 & 1.43  \\ \cline{1-5}
            
        \multirow{1}{*}{SVHN} 
            & 60.01 & 20.02 & 60.23 & 18.86 \\ \hline
            
        \multirow{1}{*}{EuroSAT} 
            & 75.45 & 8.67 & 75.44 & 1.13 \\ \hline
            
        \end{tabular}
	}
	\label{februus}
\end{table}

\begin{table}[!t]\renewcommand{\arraystretch}{1.2}

  \centering
  \fontsize{6}{8}\selectfont
  \caption{The performance of several training-phase defense methods against trojaned encoders. Since they are not applicable to the encoder, we concatenate the encoder and the downstream classifier and treat it as a classifier. }
  \subfloat[Neural Cleanse~\cite{wang2019neural} predicts a model to be trojaned if the anomaly index is larger than 2]{
        \begin{tabular}{|c|c|c|} \hline
        
        \makecell{Pre-training dataset} & \makecell{Downstream\\dataset} & \makecell{Anomaly index} \\ \hline \hline
        
        \multirow{3}{*}{CIFAR10} 
            & STL10 & 0.94     \\ \cline{2-3}
            & SVHN & 1.18     \\ \cline{2-3}
            & EuroSAT & 0.70     \\ \cline{1-3}
            
        \end{tabular}
  \label{table:nc}
  }
  
  \subfloat[MNTD~\cite{xu2019detecting} trains a meta neural classifier on a set of clean and trojaned models]{
        \begin{tabular}{|c|c|c|} \hline
        
        \makecell{Pre-training\\dataset} & \makecell{Downstream\\dataset} & \makecell{Detection accuracy on\\trojaned encoders (\%)} \\ \hline \hline
        
        \multirow{3}{*}{CIFAR10} 
            & STL10 & 53.12     \\ \cline{2-3}
            & SVHN & 50.51     \\ \cline{2-3}
            & EuroSAT & 55.67     \\ \cline{1-3}
            
        \end{tabular}
    \label{table:mntd}
    }
    
  \subfloat[I-BAU~\cite{zeng2022ibau} unlearns the trojan from the trojaned model by implicit hypergradient]{
        \begin{tabular}{|c|c|c|c|c|c|} \hline
        \multirow{2}{*}{\makecell{Pre-training\\dataset}} & \multirow{2}{*}{\makecell{Downstream\\dataset}} & \multicolumn{2}{c|}{\makecell{Trojaned encoders\\without I-BAU}} & \multicolumn{2}{c|}{\makecell{Trojaned encoders,\\wtih I-BAU}} \\ \cline{3-6} 
        
        && \makecell{ACC (\%)} & \makecell{ASR (\%)} & \makecell{ACC (\%)} & \makecell{ASR (\%)}  \\ \hline \hline
        
        \multirow{3}{*}{CIFAR10} 
            & STL10 & 73.75 & 100.0 & 74.16 & 95.53  \\ \cline{2-6}
            & SVHN & 60.29 & 99.98 & 62.52 & 83.28  \\ \cline{2-6}
            & EuroSAT & 75.59 & 100.0 & 72.50 & 96.00  \\ \cline{1-6}
            
        \end{tabular}
    \label{table:i_bau}
    }
\label{table:other_defenses}
\end{table}

\begin{table}[!t]\renewcommand{\arraystretch}{1.2}
  \centering
  \fontsize{5}{8}\selectfont
  \caption{Performance of {\name} in defending against trojan attacks to 2 real-world encoders on larger-scale datasets, ImageNet100~\cite{tian2019contrastive} and Caltech101~\cite{li2006caltech101}. The attack targets are ``stop sign'' and ``hummingbird'', respectively. }
  \subfloat[Trojan detection]{
        \begin{tabular}{|c|c|c|c|} \hline
        
        \makecell{Pre-training\\dataset} & \makecell{Downstream\\dataset} & \makecell{FPR (\%)} & \makecell{FNR (\%)} \\ \hline \hline
        
        \multirow{2}{*}{ImageNet} 
            & ImageNet100 & 4.15 & 0.57     \\ \cline{2-4}
            & Caltech101  & 1.26 & 0.05     \\ \cline{1-4}

        \multirow{2}{*}{CLIP} 
            & ImageNet100 & 2.96 & 0.24     \\ \cline{2-4}
            & Caltech101  & 1.27 & 0.98     \\ \cline{1-4}
            
        \end{tabular}
  \label{table:case_study_detection_larger}
  }

  \subfloat[Trojan removal]{
        \begin{tabular}{|c|c|c|c|c|c|} \hline
        \multirow{2}{*}{\makecell{Pre-training\\dataset}} & \multirow{2}{*}{\makecell{Downstream\\dataset}} & \multicolumn{2}{c|}{\makecell{Attacked encoders\\without {\name}}} & \multicolumn{2}{c|}{\makecell{Attacked encoders,\\wtih {\name}}} \\ \cline{3-6} 
        
        && \makecell{ACC (\%)} & \makecell{ASR (\%)} & \makecell{ACC (\%)} & \makecell{ASR (\%)}  \\ \hline \hline
        
        \multirow{2}{*}{ImageNet} 
            & ImageNet100 & 81.05 & 97.75 & 80.73 & 9.93  \\ \cline{2-6}
            & Caltech101  & 75.34 & 86.37 & 75.58 & 5.82  \\ \cline{1-6}
            
        \multirow{2}{*}{CLIP} 
            & ImageNet100 & 85.90 & 96.42 & 85.83 & 2.49  \\ \cline{2-6}
            & Caltech101  & 81.63 & 95.33 & 80.62 & 1.47  \\ \cline{1-6}
            
        \end{tabular}
    \label{table:case_study_removal_larger}
    }
\label{table:case_study_larger}
\end{table}

\begin{table}[!t]\renewcommand{\arraystretch}{1.2}

  \centering
  \fontsize{6}{8}\selectfont
  \caption{The performance of our {\name} in detecting and removing the trojan from the test inputs for a larger trigger.} 
  \subfloat[Trojan detection]{
        \begin{tabular}{|c|c|c|} \hline
        
        \makecell{Downstream\\dataset} & \makecell{FPR (\%)} & \makecell{FNR (\%)} \\ \hline \hline
        
        STL10 & 0.85 & 1.03     \\ \cline{1-3}
        SVHN & 0.14 & 1.56     \\ \cline{1-3}
        EuroSAT & 0.11 & 4.44     \\ \cline{1-3}
            
        \end{tabular}
  \label{table:detecting_larger_trigger}
  }
  
  \subfloat[Trojan removal]{
        \begin{tabular}{|c|c|c|c|c|} \hline
        \multirow{2}{*}{\makecell{Downstream\\dataset}} & \multicolumn{2}{c|}{\makecell{Trojaned encoders\\without {\name}}} & \multicolumn{2}{c|}{\makecell{Trojaned encoders,\\wtih {\name}}} \\ \cline{2-5} 
        
        & \makecell{ACC (\%)} & \makecell{ASR (\%)} & \makecell{ACC (\%)} & \makecell{ASR (\%)}  \\ \hline \hline
        
        \multirow{1}{*}{STL10} 
            & 74.00 & 100.0 & 73.53 & 4.39  \\ \cline{1-5}
            
        \multirow{1}{*}{SVHN} 
            & 60.71 & 100.0 & 59.67 & 18.25 \\ \hline
            
        \multirow{1}{*}{EuroSAT} 
            & 75.41 & 100.0 & 75.33 & 13.44 \\ \hline
            
        \end{tabular}
    \label{table:removing_larger_trigger}
    }
\label{table:larger_trigger}
\end{table}

\begin{table}[!t]\renewcommand{\arraystretch}{1.2}

  \centering
  \fontsize{6}{9}\selectfont
  \caption{The attack performance of the adapted blending trojan attacks to self-supervised learning encoders.}
        \begin{tabular}{|c|c|c|c|c|} \hline
        
        \multirow{2}{*}{\makecell{Downstream\\dataset}} & \multicolumn{2}{c|}{\makecell{Clean encoders}} & \multicolumn{2}{c|}{\makecell{Trojaned encoders}} \\ \cline{2-5}
        
        & \makecell{ACC (\%)} & \makecell{ASR (\%)} & \makecell{ACC (\%)} & \makecell{ASR (\%)} \\ \hline \hline
        
        \multirow{1}{*}{STL10} 
            & 74.38 & 1.01 & 61.40 & 68.70     \\ \cline{1-5}
            
             \multirow{1}{*}{SVHN} 
            & 54.13 & 23.44 & 60.76 & 0.00     \\ \cline{1-5}
            
             \multirow{1}{*}{EuroSAT} 
            & 74.41 & 3.52 & 72.04 & 100.0     \\ \hline
            
        \end{tabular}
\label{blending}
\end{table}

\begin{figure}[!t]
	 \centering
\subfloat[Trojan detection]{\includegraphics[width=0.23\textwidth]{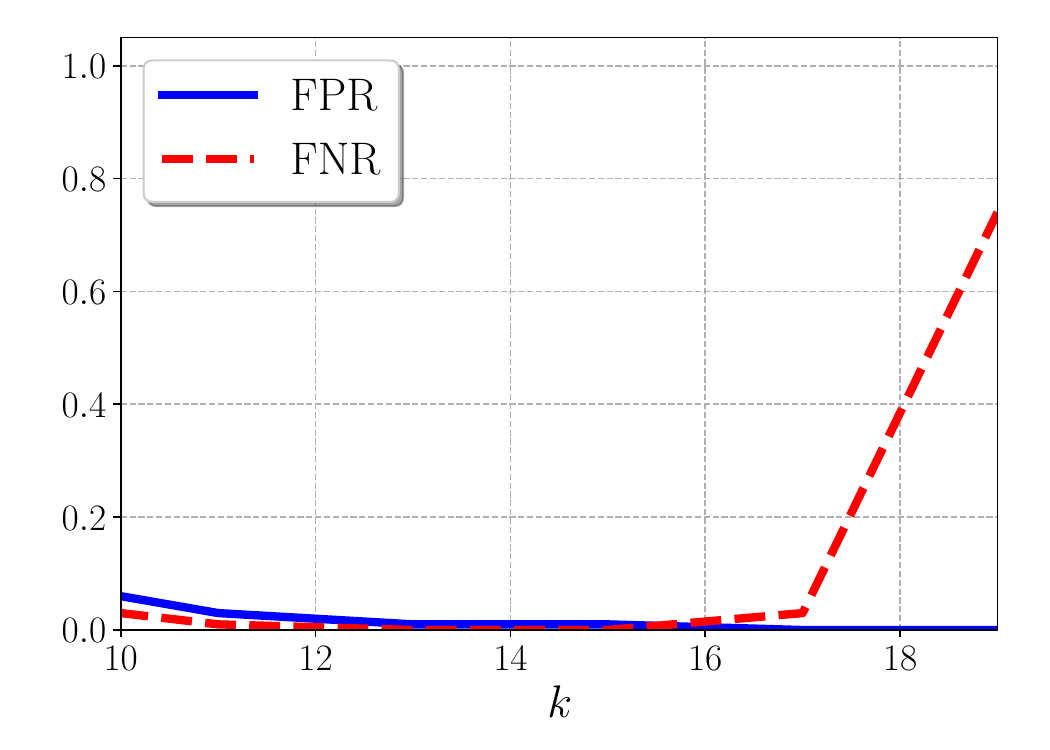}}
\subfloat[Trojan removal]{\includegraphics[width=0.23\textwidth]{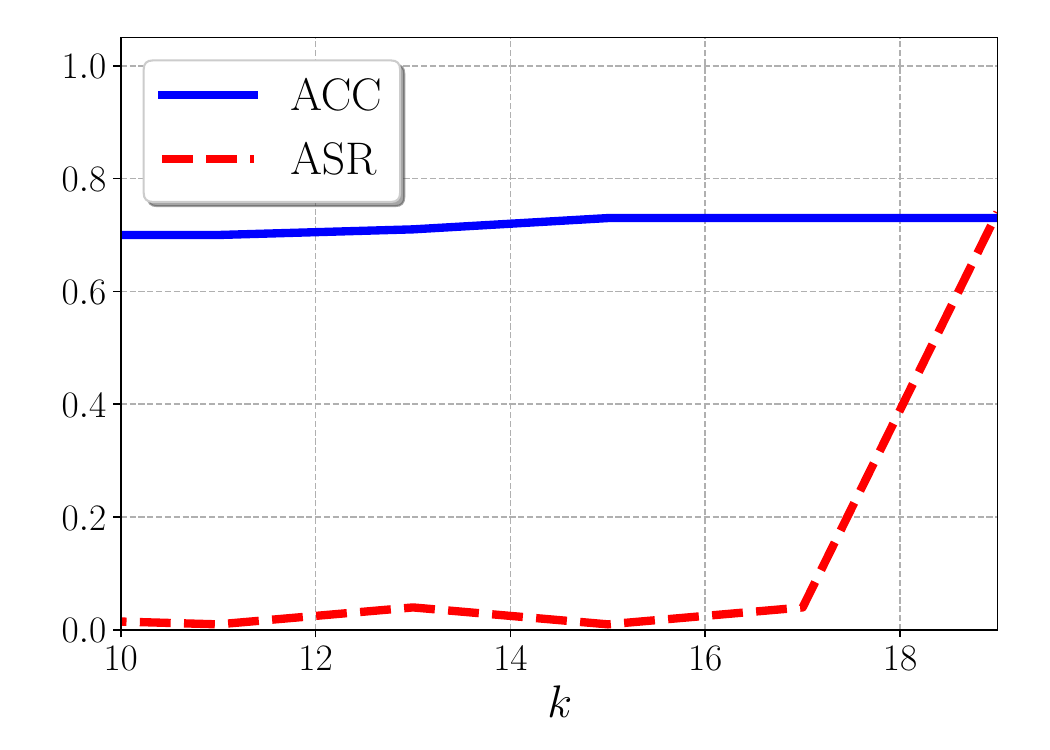}}
\caption{Impact of $k$.}
\label{impact_of_k}
\end{figure}

\begin{figure}[!t]
	 \centering
\subfloat[Trojan detection]{\includegraphics[width=0.23\textwidth]{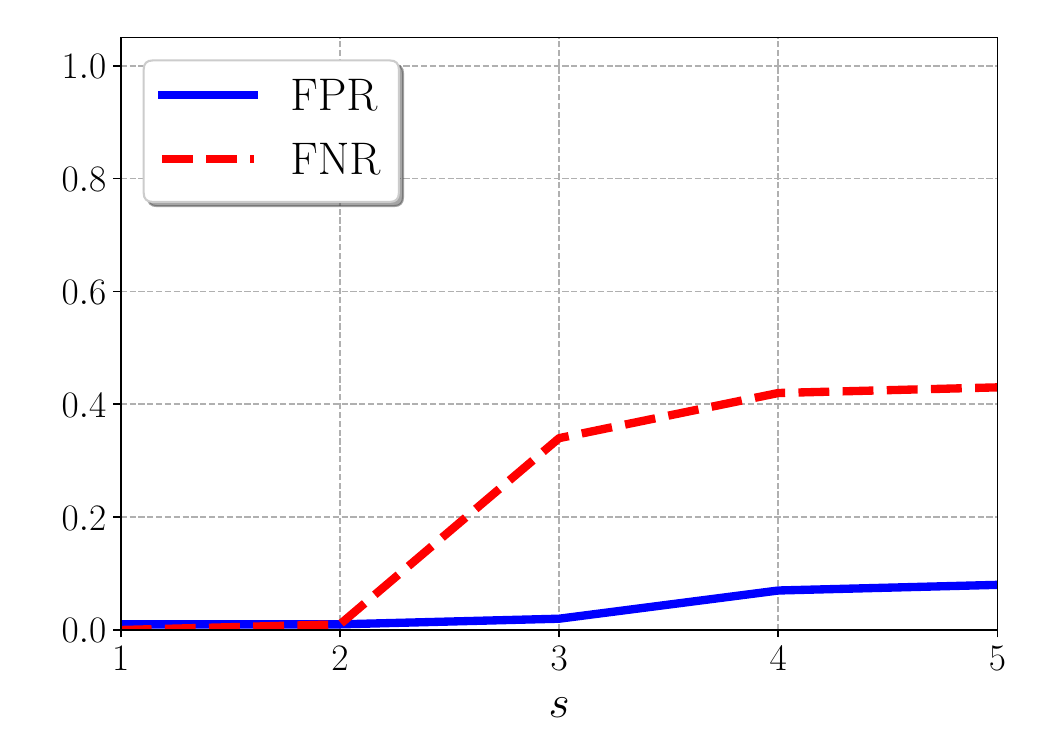}}
\subfloat[Trojan removal]{\includegraphics[width=0.23\textwidth]{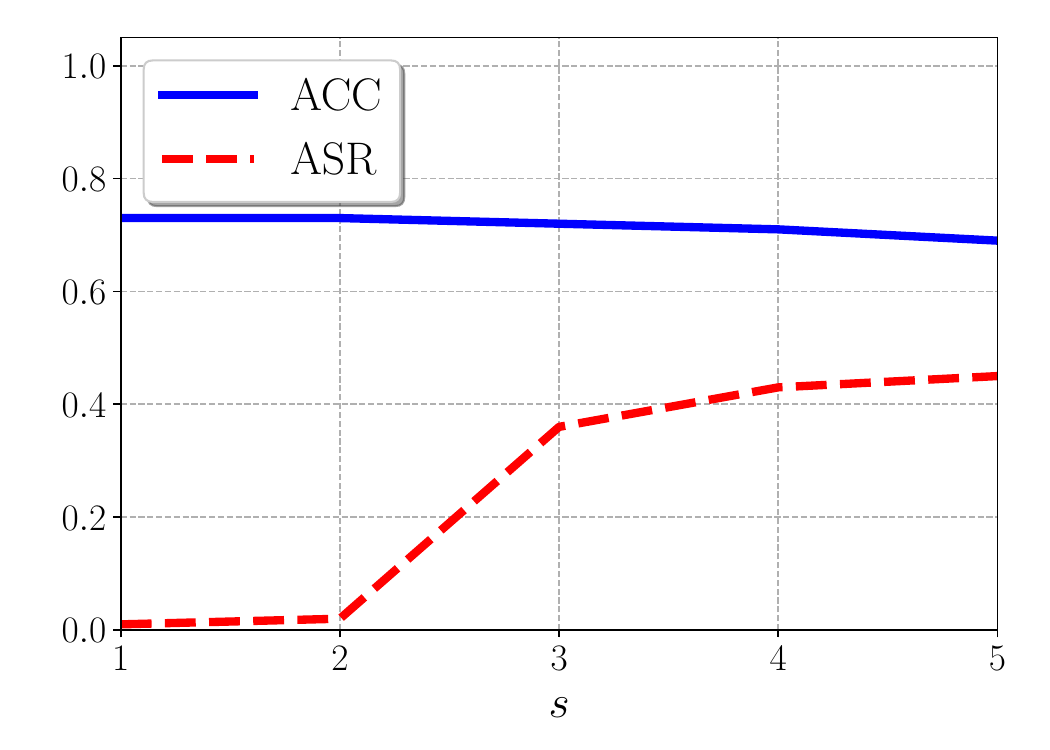}}
\caption{Impact of $s$.}
\label{impact_of_s}
\end{figure}

\begin{table}[!t]\renewcommand{\arraystretch}{1.2}

  \centering
  \fontsize{6}{8}\selectfont
  \caption{The performance of our {\name} in detecting and removing the trojan from the test inputs for other trojan attacks.} 
  \subfloat[Trojan detection]{
        \begin{tabular}{|c|c|c|} \hline
        
        \makecell{Attack method} & \makecell{FPR (\%)} & \makecell{FNR (\%)} \\ \hline \hline
        
        \multirow{1}{*}{Saha et al.~\cite{saha2022backdoor}} 
            & 2.33 & 0.00     \\ \cline{1-3}
            
            \multirow{1}{*}{Liu et al.~\cite{liu2022poisonencoder}} 
            & 1.00 & 1.03     \\ \hline
            
        \end{tabular}
  \label{table:other_attack_detection}
  }
  
  \subfloat[Trojan removal]{
        \begin{tabular}{|c|c|c|c|c|} \hline
        \multirow{2}{*}{\makecell{Attack method}} & \multicolumn{2}{c|}{\makecell{Trojaned encoders\\without {\name}}} & \multicolumn{2}{c|}{\makecell{Trojaned encoders,\\wtih {\name}}} \\ \cline{2-5} 
        
        & \makecell{ACC (\%)} & \makecell{ASR (\%)} & \makecell{ACC (\%)} & \makecell{ASR (\%)}  \\ \hline \hline
        
        \multirow{1}{*}{Saha et al.~\cite{saha2022backdoor}} 
            & 73.64 & 89.72 & 72.49 & 1.32  \\ \cline{1-5}
            
        \multirow{1}{*}{Liu et al.~\cite{liu2022poisonencoder}} 
            & 73.04 & 90.03 & 72.60 & 4.93 \\ \hline
            
        \end{tabular}
    \label{table:other_attack_removal}
    }
\label{table:other_attack}
\end{table}

\end{document}